\documentclass[12pt]{article}
\usepackage[utf8]{inputenc}
\usepackage{amsmath,amsfonts,amssymb,amsthm,bm,cite,fullpage,graphicx,color}
\usepackage{subcaption}
\usepackage{subfig}
\title{Entanglement in N-harmonium: bosons and fermions}

\author{C. L. Benavides-Riveros$^{1,2}$, I. V. Toranzo$^{3}$,
and J. S. Dehesa$^{3}$
\\ \\
$^1$Departamento de F\'isica Te\'orica,
\\
Universidad de Zaragoza, 50009 Zaragoza, Spain
\\ \\
$^2$Instituto de Biocomputaci\'on  y F\'isica de Sistemas Complejos,
\\
Mariano Esquillor (Edificio I+D), 50018 Zaragoza, Spain
\\ \\
  $^3$Instituto Carlos I de F\'{\i}sica Te\'orica y Computacional 
  \\
  and Departamento de F\'{\i}sica At\'omica Molecular
y Nuclear, 
\\
Universidad de Granada, 18071 Granada, Spain.
}
  
\date{\today}

\DeclareMathOperator{\Tr}{Tr}         %% trace

\newcommand{\al}{\alpha}            %% short for \alpha
\newcommand{\bt}{\beta}             %% short for \beta
\newcommand{\Dl}{\Delta}            %% \Delta
\newcommand{\dl}{\delta}            %% \delta
\newcommand{\Ga}{\Gamma}            %% \Gamma
\newcommand{\ga}{\gamma}            %% \gamma
\newcommand{\la}{\lambda}           %% \lambda
\newcommand{\om}{\omega}            %% short for \omega
\newcommand{\vf}{\varphi}           %% \varphi
\newcommand{\vs}{\varsigma}         %% \varsigma
\newcommand{\A}{\mathcal{A}}        %% functional
\newcommand{\V}{\mathcal{V}}        %% functional
\newcommand{\C}{\mathbb{C}}         %% complex numbers
\newcommand{\dn}{{\mathord{\downarrow}}} %% `down' spinors
\newcommand{\half}{\tfrac{1}{2}}    %% small fraction  1/2
\newcommand{\twobytwo}[4]{\begin{pmatrix}#1& #2\\ #3& #4\end{pmatrix}}
\newcommand{\up}{{\mathord{\uparrow}}} %% `up' spinors
\newcommand{\word}[1]{\quad\mbox{#1}\quad} %% well-spaced words
\newcommand{\x}{\times}             %% cartesian product
\renewcommand{\.}{\cdot}            %% scalar product

\newcommand{\gs}{\mathrm{gs}}       %% ground state
\newcommand{\vecform}{\bm}              %% format for trivectors
\newcommand{\pp}{\vecform{p}}           %% trivector p
\newcommand{\rr}{\vecform{r}}           %% trivector r
\newcommand{\vxi}{\vecform{\xi}}           %% trivector xi
\newcommand{\vXi}{\vecform{\Xi}}           %% trivector Xi
\newcommand{\uu}{\vecform{u}}           %% trivector u
\newcommand{\xx}{\vecform{x}}           %% trivector x
\newcommand{\yy}{\vecform{y}}           %% trivector y
\makeatletter
\def\section{\@startsection{section}{1}{\z@}{-3.5ex plus -1ex minus
 -.2ex}{2.3ex plus .2ex}{\large\bfseries}}
\def\subsection{\@startsection{subsection}{2}{\z@}{-3.25ex plus -1ex
 minus -.2ex}{1.5ex plus .2ex}{\normalsize\bfseries}}
\makeatother

\begin{document}

\maketitle

\begin{abstract}
The ground-state entanglement of a single particle of the N-harmonium system 
(i.e., a completely-integrable model of $N$ particles where both the confinement 
and the two-particle interaction are harmonic) is shown to be analytically determined 
in terms of $N$ and the relative interaction strength. 
For bosons, we compute the von Neumann entropy of the one-body reduced 
density matrix by using the corresponding natural occupation numbers. 
There exists a critical number $N_c$ of particles so that below it, for positive values 
of the coupling constant, the entanglement grows when the 
number of particles is increasing; the opposite occurs for 
$N > N_c$. For fermions, we compute the one-body reduced density
matrix for the closed-shell spinned case. In the strong coupling regime, the linear entropy 
of the system decreases when $N$ is growing. For fixed $N$, the entanglement is 
found (a) to decrease (increase) for negatively (positively) increasing values of the 
coupling constant, and (b) to grow when the energy is increasing. Moreover, the spatial and spin contributions to the total entanglement are found to be of comparable size.
\end{abstract}

% 31.15.-p Calculation of electronic structure of Atoms
% 03.67.-a Quantum information
%05.30.Fk Fermion systems and electron gas

\newpage 

\section{Introduction}
\label{sec:introibo}

The most precise determination of the properties of finite many-electron systems 
is usually done by means of a full-configuration-interaction (FCI) method, where 
the solution of the corresponding Schr\"odinger equation in a given one-electron basis 
is expressed in terms of a linear combination of all possible Slater determinants. 
Its application is naturally reduced to a bunch of small systems because of the 
enormous number of the involved determinants \cite{Szabo, Helgaker}. 

Let us highlight that for all FCI approaches the correlation effects, which remain 
solely in the wave function, are not described by any correlation operator. Moreover, 
these effects are usually numerically computed as a difference of two variational 
energy bounds. Thus, up to now the electron correlation is widely and implicitly 
believed to be a purely methodical effect coming from the inadequate use of a 
trial wave function of multiconfiguration Hartree-Fock type; so, lacking of physical reality.

The application of quantum information ideas and techniques in electronic structure 
theory has recently allowed to conclude that the electron correlation is closely related 
to entanglement of electrons. Indeed, it has been proved that while the single Slater 
determinant in the monoconfigurational Hartree-Fock approximation is a disentangled 
state, the wavefunction of the multiconfiguration Hartree-Fock approximations (such as, 
e.g. FCI) accounts for entanglement effects. Therefore, entanglement plays an essential 
role not only in quantum communication between parties separated by macroscopic 
distances (see e.g., \cite{NIELSEN, HOR}), but also it is essential to characterize 
quantum correlations at short distances. The latter problem, where one should 
necessarily consider the indistinguishable character of the involved particles, has received 
relatively less attention until a short time ago \cite{CIraciamici, HUAN, WAN, Marzo, Marzo1,TMB10}. 
This is a serious lack because of its relevance for quantum information processing 
in various physical systems (see e.g., \cite{Amico, CIraciamici}), to gain deeper 
insight into non-classical correlations of atomic and molecular systems as well 
as to fully understand the course of their dissociation processes and chemical 
reactions \cite{TMB10, Amico, ESQ0}.

The main difficulty, however, stems from the fact that the Schr\"odinger equation 
of most quantum many-body systems cannot be solved analytically. Even numerically, 
the determination of the wavefunction is, in general, a serious problem. In the last 
three years an intense effort has been made to determine the entanglement of some 
real atomic and molecular species such as helium-like atoms \cite{MANZA, Dehesa, 
Benenti, HOFER, LIN}, of a few processes of diatomic molecules \cite{ESQ1} and 
elementary chemical reactions \cite{ESQ2}. These works basically focus on the 
entanglement of bipartite systems, mainly because the characterization of this 
phenomenon for systems of many indistinguishable constituents is much less known, 
even at the level of the very notion of entanglement measure \cite{magyares}.

Thus, the quantification of entanglement of bound states for model systems 
enabling analytic solutions of the associated Schrödinger equation is being a 
promising way to investigate correlation phenomena. Indeed, entanglement 
between the constituents of any bound system is most conveniently analyzed 
in such models, enabling to relate it to the bosonic or fermionic character. Up 
to now, however, only entanglement of some models of two bound electrons 
have been determined. We refer to the 2-harmonium (or Moshinsky) \cite{HEI, 
Moshinsky1, NagyP09, YAN, Laetitia}, Crandall and Hooke\cite{MANZA} atoms. 
In all these models the electron confinement is harmonic, and the electron-electron 
interaction is of harmonic (2-harmonium), $r_{12}^{-1}$ (Crandall), and Coulombic 
(Hooke) type. All of them show that when the spin degree of freedom and the
indistinguishability of electrons are taken into account, new entanglement aspects\cite{YAN} 
are encountered as compared to the model of distinguishable particles, although 
some further clarification is needed.

It is also worth noting that some entanglement features of the 2-harmonium atom 
are also qualitatively reproduced by the other two models, which give a good 
description for certain two-fermion systems (e.g., two electrons in a quantum 
dot is approximately described by the Hooke model). Namely, the growth of the 
entanglement when either the relative strength or the excitation energy is increasing. 
It is most interesting that recently the entanglement of some real helium-like atoms has 
been numerically shown to have an increasing dependence on the energy. This has 
been done both by the use of very accurate one-electron basis functions of Hylleraas-Kinoshita 
type \cite{Dehesa, LIN} and some Gaussian or Slater type orbital basis 
sets \cite{HUAN, Benenti}. As well it is observed from the models that the entanglement 
decrease of these systems in terms of the nuclear charge Z can also be understood 
as the result of the relative decrease of the electron-electron-interaction.
  
In this work we will study the entanglement of the one-body reduced density 
matrix of the N-harmonium model for bosons and fermions analytically. 
Von Neumann (for bosons) and linear (for fermions) entropies will be used. 
The one-body reduced matrix for the spinless fermionic case has been previously 
computed \cite{Schilling}. We obtain for the first time an explicit expression of the
 one-body reduced density matrix for the \textit{closed-shell} spinned fermionic-case.
 This study will allow us to show that some entanglement features of finite many-par\-ticle 
 systems can be understood by purely kinematical considerations to a certain extent. 
 The N-particle harmonium is a completely-integrable system with an arbitrary 
 number $N$ of particles where both the confinement and the two-particle interaction 
 are harmonic. This model has also been used to study cold atoms \cite{Zinner} and 
 to gain further insight into numerous phenomena of a variety of physical systems up 
 to black holes (see e.g.,\cite{HEI, Bombelli, Johnson, Srednicki, Zinner2, Koscik, 
 LosChinos, ETH}).

The paper is structured as follows. First, in Section \ref{sec:atom}
we briefly formulate the quantum-mechanical problem of the N-particle harmonium, 
showing its separability by using the appropriate set of normal coordinates 
\cite{Moshinskyreturns}, and fixing our notational settings. Then, in Section 
\ref{sec:bosones} we discuss the general mathematical structure of the one-body 
reduced density matrix in the bosonic case, obtaining its explicit expression. Moreover, 
we compute and discuss the analytical expression of the von Neumann entropy 
of the one-body reduced density matrix for the N-boson harmonium. Further, in Sections \ref{sec:fermions} and \ref{sec:spin} 
we obtain explicit expressions for the one-body reduced density matrix of the spinned
N-fermion harmonium and analyze the linear entropy of the one-body reduced density 
matrix, with and without the spin degree of freedom. Finally, some concluding 
remarks and two appendices are given.

% \S 2
\section{The N-harmonium problem}
\label{sec:atom}

The N-harmonium model is a system of $N$ interacting particles (fermions or 
bosons) which interact harmonically in a three-dimensional 
harmonic well. It is characterized by the Hamiltonian 
\begin{equation}
H = \frac12 \sum^N_{i=1}|\pp_i|^2 + \frac{k}{2} \sum^N_{i=1}|\rr_i|^2
+ \frac{\dl}{2} \sum^N_{i\,<\,j} r_{ij}^2,
\label{eq:Mosh-atom}
\end{equation}
where $r_{ij} := |\rr_i-\rr_j|$, $k$ is the coupling constant of the harmonic well 
and $\dl$ the coupling of the harmonic interaction between the particles.
The treatment of harmonically interacting bosons by means of
this Hamiltonian represents the first exact solution of a N-particle 
system using only conditions on the reduced space of two-particle density matrices \cite{Gido}.
This system can be expressed in a separable form (i.e., as a system of 
uncoupled oscillators) using the set of normal coordinates
$\{\vxi_1, ..., \vxi_N\}$ given by
\begin{align}
\vxi_N := \frac{1}{\sqrt{N}} \sum^N_{i=1} \rr_i \word{and} 
\vxi_m := \frac{1}{\sqrt{m(m+1)}} \sum^m_{i=1}(\rr_i - \rr_{m+1}),
\end{align}
with $m \in \{1,\ldots,N-1\}$.
This is an orthogonal transformation of the position variables. A similar change of coordinates 
for momenta results in a canonical transformation, preserving the symplectic form. 
Let us call the new set of momenta
 $\{\vXi_m\}$. A direct calculation shows
that 
\begin{align}
\sum^{N-1}_{m=1} \vxi_m^2 = \frac{1}{N}\sum^N_{i\,<\,j} r_{ij}^2 =
\sum^N_{m=1} \rr_m^2 -  \vxi_N^2,
\end{align}
so that the Hamiltonian \eqref{eq:Mosh-atom} can be expressed in the following separable form
\begin{align}
H = \mathcal{H}_N + \sum^{N-1}_{m=1} \mathcal{H}_m,
\word{where}
\mathcal{H}_N = \frac{1}{2} \vXi_N^2 + \frac{1}{2}\om^2 \vxi_N^2 \word{and}
\mathcal{H}_m = \frac12 \vXi_m^2 + \frac12 \mu^2 \vxi_m^2,
\end{align}
where $\om^2 := k$, and $\mu^2 := k + N\dl$ depending on the number of particles. 
Then, the physical solutions of the associated Schrödinger equation (i.e., the wave 
functions $\Psi(\xx_1,\cdots, \xx_N)$), can be readily obtained. Here, let us first note 
that (a) there is a ground state whenever $\mu^2 > 0$, and (b) the particles are no
longer bound if the relative interaction strength 
\begin{align}
\frac{\dl}{k} \le -\frac1N.
\label{eq:maximum}
\end{align}

On the other hand, let us comment here that the entanglement problem is 
formulated in terms of reduced density matrices. Either directly from the resulting 
one-particle reduced density matrix or, since the particles are assumed to interact 
pairwise, in terms of the two-particle density matrix 
\begin{align}
\rho(\xx_1,\xx_2;\yy_1,\yy_2) := 
\int d\xx_3 \cdots d\xx_N \; |\Psi(\xx_1,\xx_2,\xx_3,\cdots, \xx_N)
\rangle \langle\Psi(\yy_1,\yy_2,,\xx_3,\cdots, \xx_N)|, 
\end{align}
which carries all the necessary information required for
calculating the quantum-mechanical properties of the whole system.
The symbol $\xx$ stands for spatial and spin coordinates, $\xx~:=~(\rr,\vs)$.
In particular, the ground state energy of the system can be computed 
by minimizing  a simple linear functional of $\rho$. In passing, let us mention that the 
N-representability problem for this matrix has proved to be a major challenge for 
quantum chemistry \cite{Mazziotti}. The one-body density matrix, which is the basic 
variable in reduced density matrix functional theory \cite{Lapernal}, is then given as
\begin{align}
\rho_1(\xx;\yy) := \int
 \rho(\xx,\xx_2;\yy,\xx_2) d\xx_2.
\end{align}
By means of the spectral theorem, $\rho_1(\xx;\yy)$ can be decomposed
 in terms of its natural spin orbitals $\{\phi_i(\xx)\}$ and its eigenvalues $\{n_i\}$,
 the \textit{natural occupation numbers}: $\rho_1(\xx;\yy) = \sum_i n_i \phi_i(\xx) \phi_i(\yy)$,
where $\sum_i n_i = 1$, with (for fermions) $0 \leq n_i \leq 1/N$.

For the case of an assembly of bosons we will compute the von Neumann entropy 
of the one-body reduced density matrix
finding explicitly the occupation numbers of the system, as described in the next 
section. Recently, there has been a renewed interest in formulating the
reduced density matrix theory using Wigner quasidistributions
\cite{Laetitia, Dahl09, Pluto}. Our treatment for bosons will be done by using  
the Wigner function in phase space, what is a natural procedure for these systems. Let 
us advance that there exists a critical value for the number of particles. Below this 
critical value (around 3.5), we will show that for positive values of the coupling constant $\dl$ 
the first occupation number decreases (and consequently the von Neumann entropy grows) 
as the number of particles is increasing. Above this value, the first occupation 
number increases as the number of particles is increasing. Moreover, the first occupation 
number tends to 1 (and the entropy tends to 0) in the limit when the number of particles 
tends to infinity.

Later on, for fermions we use the one-body reduced density matrix to compute an explicit 
expression for its purity and its linear entropy as well. Let us advance that we will show 
that in the region of negative coupling $\dl$, the entropy grows when the number of particles 
is increasing. A similar situation is observed in the attractive case for small values of the coupling
constant. For large values of $\dl/k$ the situation is the opposite: when particles are added to the
system the entropy decreases. In both bosonic and fermionic cases, we will show that it is possible to calculate these two measures of entanglement as a function of the coupling constant $\dl/k$ and the 
number of particles.

% \S 3
\section{The N-boson harmonium: von Neumann entropy}
\label{sec:bosones}

In this section we determine the von Neumann entropy of the N-boson harmonium 
ground state in terms of $N$ and the relative interaction strength $\dl/k$. Taking into 
account the Gaussian character of the ground-state oscillator wave function, the 
ground state distribution on phase space is characterized by the Wigner N-body 
density function
\begin{align}
&W^{b}_\gs(\rr_1,\ldots,\rr_N;\pp_1,\ldots,\pp_N) = \frac{1}{\pi^{3N}}
\, e^{ - 2\mathcal{H}_N/\om}  e^{-2 \sum^{N-1}_{m=1}  \mathcal{H}_m/\mu} \nonumber \\
&\qquad = \frac{1}{\pi^{3N}} \, \exp\bigg[- (\om - \mu) \, \vxi^2_N -
\mu \sum^N_{i=1} \rr_i^2 + \frac{\om - \mu}{\om\mu} \, \vXi^2_N -
\frac{1}{\mu} \sum^N_{i=1} \pp_i^2\bigg].
\label{eq:wigner1}
\end{align}
The energy of the ground state is the total sum of the contributions
of each individual oscillator, i.e.
$E^b_\gs = \frac32 \big[\om + (N-1) \mu\big] $. On the other hand the one-body density 
factorizes as a product of two separable quantities, to wit:
\begin{align}
d^b_1(\rr;\pp) :=& \int
W^b_\gs(\rr,\rr_2,\ldots,\rr_N;\pp,\pp_2,\ldots,\pp_N) \, 
\prod^N_{m=2} d\rr_m d\pp_m \nonumber \\
=& \, 
e^{-\mu \rr^2 - \frac{1}{\mu} \pp^2} \, 
\Dl^{\rr}(\rr,\mu,\om,N) \, \Dl^{\pp}(\pp,\mu,\om,N) .
\label{kernel}
\end{align}
It can be shown that the $\Dl$-function fulfills the following
recursion relation:
\begin{align}
\Dl^{\rr}(\rr,\mu,\om,N) :=& \frac{1}{\pi^{3N/2}}\int \exp\bigg[- \frac{\om -
\mu}{N} \bigg(\rr + \sum^N_{m=2}\rr_m\bigg)^2 - \mu \sum^N_{m=2} \rr_m^2
\bigg] \, \prod^N_{m=2} d\rr_m \nonumber \\
=& \frac{1}{\mu_{1}^{3/2}} \frac{1}{\pi^{3(N-1)/2}} \int \exp\bigg[-
\frac{\om - \mu}{N} \frac{\mu}{\mu_{1}} \bigg(\rr +
\sum^{N-1}_{m=2}\rr_m\bigg)^2 - \mu \sum^{N-1}_{m=2} \rr_m^2 \bigg] \,
\prod^{N-1}_{m=2} d\rr_m \nonumber \\
=& \frac{1}{\mu_{1}^{3/2}} \frac{1}{\pi^{3(N-1)/2}} \int \exp\bigg[-
\frac{\om - \mu_1}{N-1} \frac{\mu}{\mu_{1}} \bigg(\rr +
\sum^{N-1}_{m=2}\rr_m\bigg)^2 - \mu \sum^{N-1}_{m=2} \rr_m^2 \bigg] \,
\prod^{N-1}_{m=2} d\rr_m \nonumber \\
=& \,  ({\rm const.}) \, \Dl^{\rr}\bigg(\sqrt{\frac{\mu}{\mu_1}}
\rr,\mu_1,\om,N-1\bigg) ,
\label{eq:unodensidad}
\end{align}
where $\mu_m := \frac{m}N \, \om + \frac{N-m}{N} \mu$. The constant $({\rm const.})$ will be 
determined by normalization so that it will depend on the number of particles as well 
as on the frequencies. Therefore, 
\begin{align}
\Dl^{\rr}(\rr,\mu,\om,N) &= ({\rm const.}) \,
\Dl^{\rr}\Bigg(\sqrt{\frac{\mu}{\mu_1}} \rr,\mu_1,\om,N-1\Bigg) 
= ({\rm const.}) \,
\Dl^{\rr}\Bigg(\sqrt{\frac{\mu}{\mu_2}} \rr,\mu_2,\om,N-2\Bigg) \nonumber \\
&= \ldots
= ({\rm const.}) \,
\exp\Bigg(-\frac{\om-\mu}{N} \frac{\mu}{\mu_{N-1}} \rr^2\Bigg),
\end{align}
and in addition that 
\begin{align}
\Dl^{\pp}(\pp,\mu,\om,N) \equiv \Dl^{\rr}(\pp,\mu^{-1},\om^{-1},N),
\end{align}
so that the expression \eqref{kernel} can be transformed as:
\begin{align}
d^b_1(\rr;\pp) = \frac{N^3}{\pi^3} 
\bigg(\frac{\om\mu}{\A_N}\bigg)^{3/2}
\exp\bigg( -\frac{N\om\mu}{(N-1)\,\om + \mu} \, \rr^2 - 
\frac{N}{\om + (N-1)\,\mu} \, \pp^2\bigg),
\label{eq:quasi}
\end{align}
where $\A_N = [(N-1)\, \om +\mu] [\om + (N-1)\,\mu]$.

Let us now diagonalize this expression by decomposing it into its set
of natural orbitals as well as into its occupation numbers.  The task
is made easier by noting that the one-body quasidensity is a linear
combination of Laguerre polynomials (the phase-space counterpart of the
Hermite polynomials).  To see that, let us define
\begin{align}
\ga_N &:= (\om\mu)^{1/4} \bigg[\frac{\om+(N-1)\, \mu}{(N-1)\,\om +
\mu}\bigg]^{1/4} , \qquad \la_N := \frac{N\sqrt{\om\mu}}{\sqrt{\A_N}},
\label{mostretes1}
\end{align}
as well as the symplectic transformation
\begin{align}
S_N := \begin{pmatrix}
\ga_N & 0 \\
  0 & \ga_N^{-1}  \end{pmatrix}
 \word{and} U := S_N \uu \word{where} \uu^T = (\rr,\pp).
 \label{mostretes2}
\end{align}
Then the one-body quasidensity function can be rewritten in the Gaussian way as follows, 
\begin{align}
d^b_1(U) = \frac{\la_N^3}{\pi^3}
e^{ -\la_N U^2},
\label{eq:gaussian}
\end{align}
showing that it is actually a \textit{Gibbs state}. Note that $\la_N \le 1$ 
for all $N$, since $\om + \mu \ge 2\sqrt{\om\mu}$ and 
$\A_N \ge  N^2 \om\mu$ and that $\la_N \rightarrow 1$ when $N \rightarrow \infty$.

It is known that associated to any symplectic transformation (say,
$S_N$), there is a unitary operator acting on the Hilbert space
\cite{PequenoJuan}. Let us use this transformation to find the set of
occupation numbers in the basis of Wigner eigenfunctions of the
harmonic oscillator.  Since the one-body quasidensity function factorizes
completely, from now on we work in one dimension.  From the series
formula
\begin{align}
(1-t)\sum^{\infty}_{r=0} L_r(x) \, e^{-x/2} \, t^r = e^{-\frac{1+t}{1-t} 
\frac{x}{2}},
\end{align}
where $L_r(x)$ is the Laguerre polynomial, it follows that
\begin{align}
d^b_1(U) = \sum^{\infty}_{r=0} \frac{(-1)^r}{\pi} \, L_r(2U^2) \,
e^{-U^2} \, n_r,
\end{align}
where the occupation numbers are equal to 
\begin{align}
n_r = \frac{2\la_N}{1+\la_N} \bigg(\frac{1-\la_N}{1+\la_N}\bigg)^r
=: (1 - t_N) \,  t_N^r,
\label{eq:ON}
\end{align}
fulfilling $\sum_r n_r = 1$ as one should expect.

\begin{figure}[ht] 
\centering
\begin{subfigure}[b]{.48\textwidth}
 \centering
 \includegraphics[width=4cm]{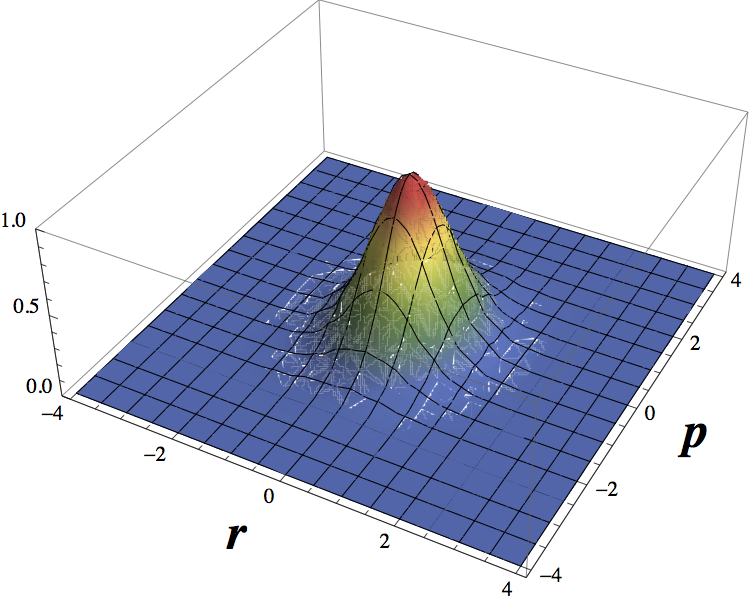} 
\caption{$N = 2$} 
\end{subfigure}
\hfill
\begin{subfigure}[b]{.48\textwidth}
\centering
\includegraphics[width=4cm]{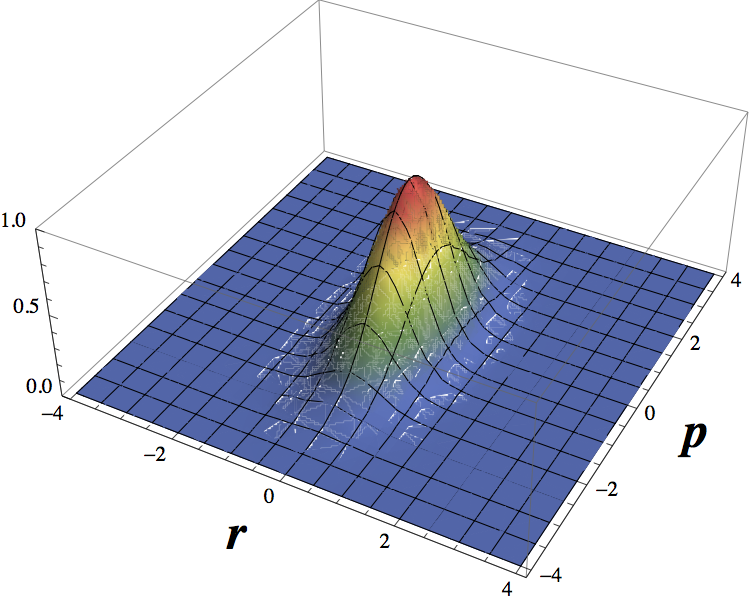}  
\caption{$N = 6$} 
\end{subfigure}
\hfill
\begin{subfigure}[b]{.48\textwidth}
\centering
\includegraphics[width=4cm]{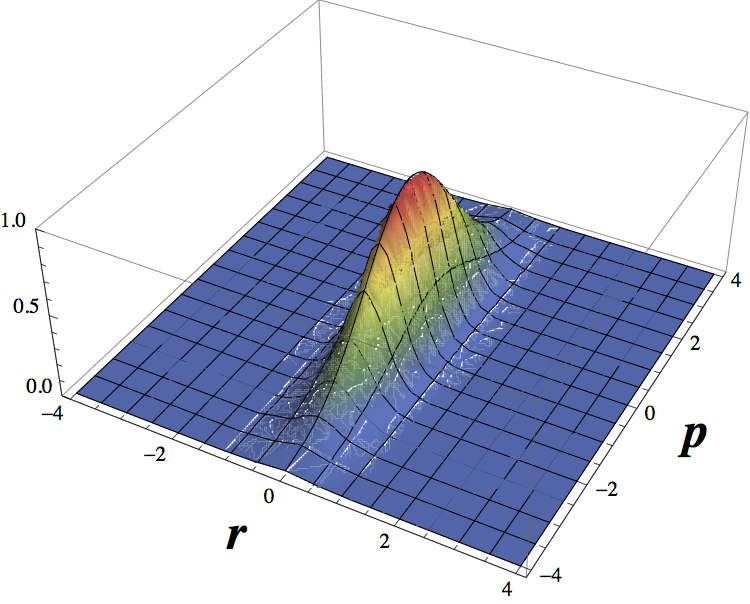}  
\caption{$N = 20$} 
\end{subfigure}
\hfill
\begin{subfigure}[b]{.48\textwidth}
\centering
\includegraphics[width=4cm]{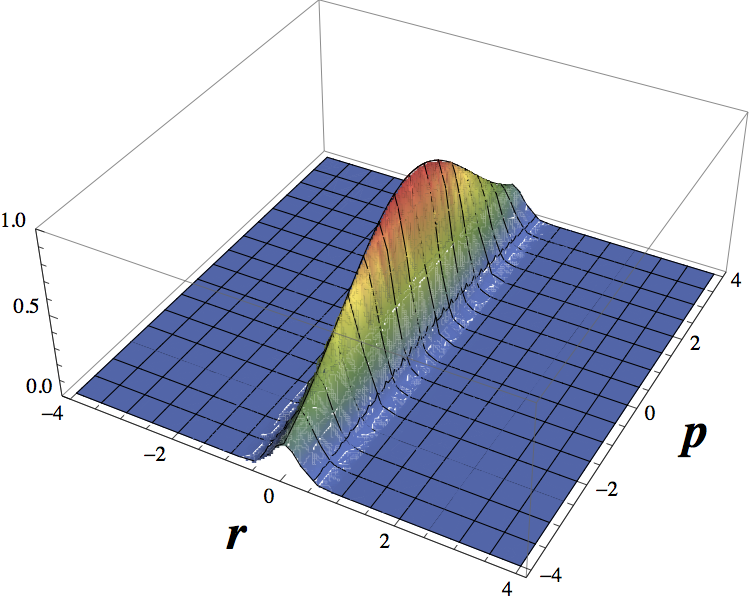}  
\caption{$N = 50$} 
\end{subfigure}
\hfill \caption{The one-body quasidensity function \eqref{eq:quasi} is plotted 
for different values of the number of bosons. Since the one-body quasidensity 
function factorizes completely, we plot it in one spatial dimension 
(two dimensions in phase space). The strength $\dl/k$ is taken to be equal to $1$. 
Note that as the number of bosons grows, the profiles of the 
position and momentum densities become narrower and wider, respectively.}
\label{graf:Wigner}
\end{figure}

For the sake of completeness we plot the numerical behavior of the one-body 
quasidensity function of the N-boson harmonium in Figure \ref{graf:Wigner} for 
different values of the number of bosons. We observe that, as the number of 
bosons grows, the profiles of the 
position and momentum densities become narrower and wider, respectively. This 
clearly indicates that the more precisely the particles
are localized in position space, the larger the localization in momentum space, 
as one should expect according to the position-momentum uncertainty principle.
There is no relevant difference for negative values of the coupling 
constant $\dl$ except that in this case it can be plotted only when $\dl/k \ge -1/N$.

Finally we can compute the von Neumann entropy of the N-boson harmonium explicitly 
in terms of $N$ and the relative interaction strength $\dl/k$, obtaining the value
\begin{align}
S(N) :=& - \sum_{r=0} n_r \, \log n_r = - \log (1 - t_N) - \frac{t_N \, 
\log(1-t_N)}{1 - t_N} \nonumber \\
=& - \frac{\sqrt{\A_N} +
N\sqrt{\om\mu}}{2N\sqrt{\om\mu}} \log\bigg[\frac{2N\sqrt{\om\mu}}{
\sqrt{\A_N} + N\sqrt{\om\mu}} \bigg],
\label{eq:vNentropy}
\end{align}
which complements and extends a similar formula encountered by other means in a 
black-hole context\cite{Srednicki}.

\begin{figure}[ht] 
\centering
\includegraphics[width=8cm]{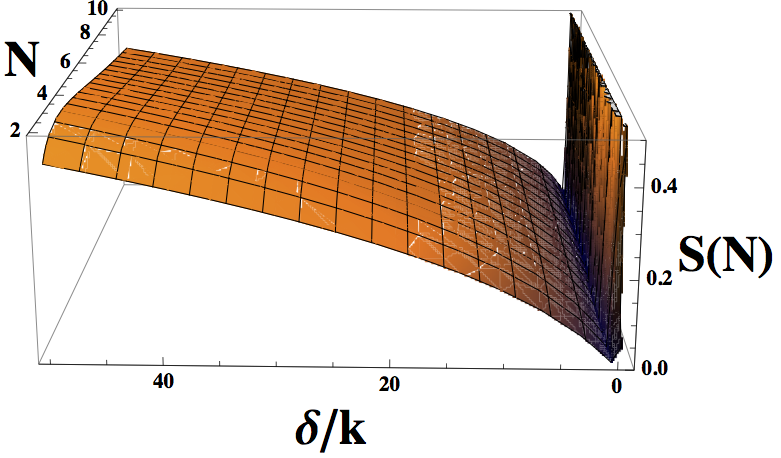} 
\caption{Von Neumann entropy of the one-body reduced density matrix
for the N-boson harmonium is plotted as
a function of the number of bosons $N$ and the relative interaction strength $\dl/k$. It is
apparent that there is a critical point around $N_c \sim 3.5$, 
where the entanglement acquires its maximum value.}
\label{graf:third}
\end{figure}

\begin{figure}[ht] 
\centering
\begin{subfigure}[b]{.48\textwidth}
 \centering
 \includegraphics[width=8.2cm]{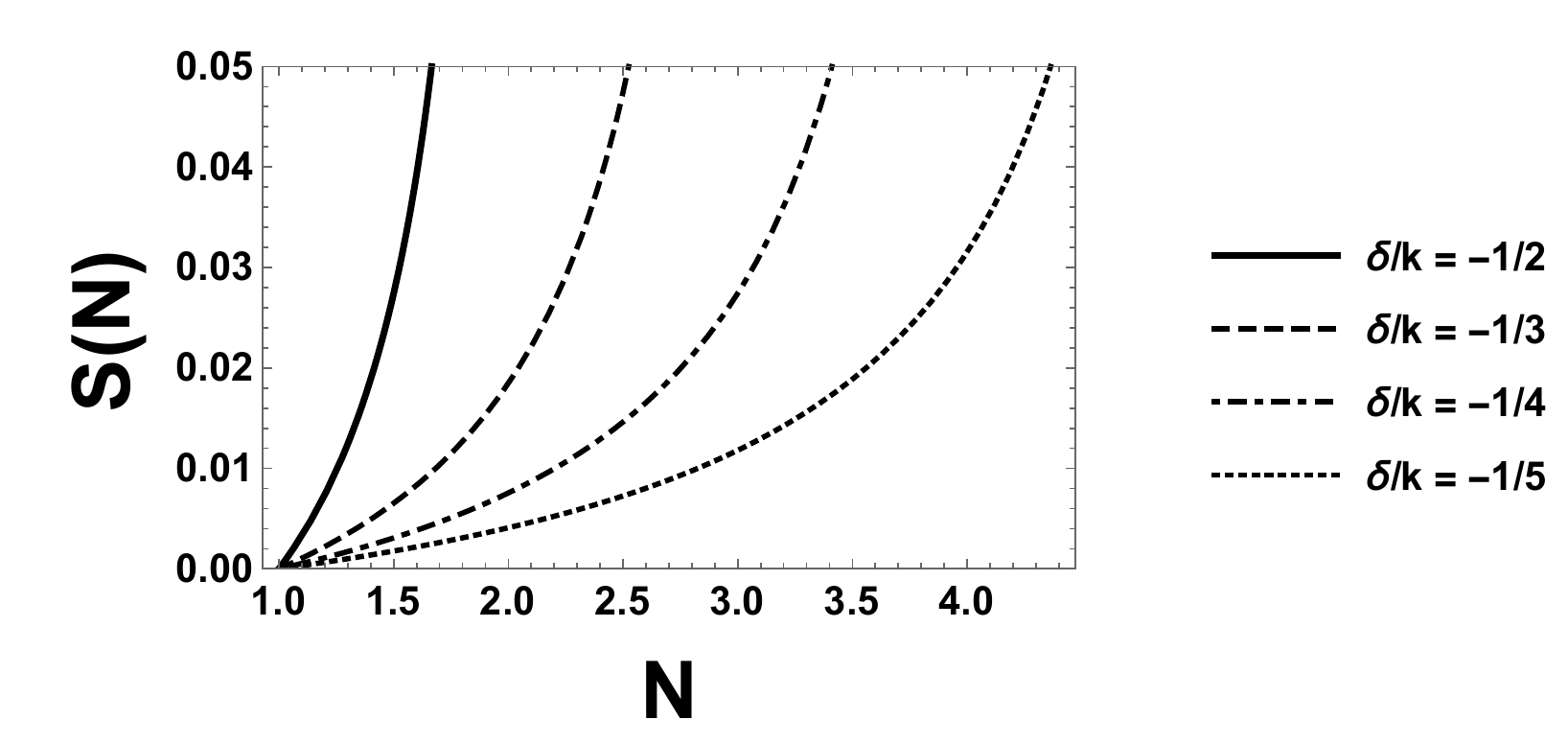} 
\end{subfigure}
\hfill
\begin{subfigure}[b]{.48\textwidth}
\centering
\includegraphics[width=8.2cm]{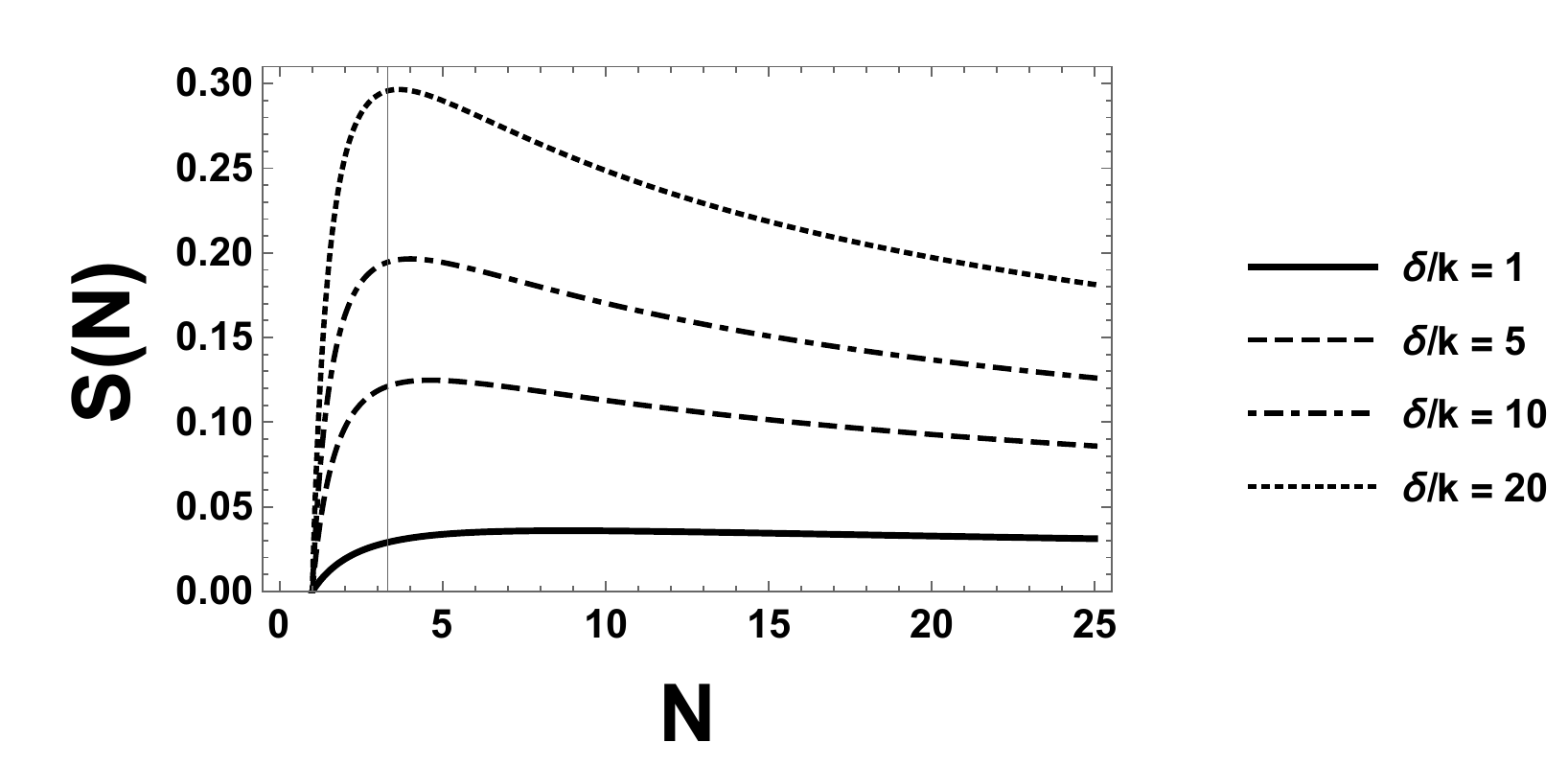}  
\end{subfigure}
\hfill
\caption{Von Neumann entropy of the one-body reduced density matrix
for the N-boson harmonium, whose explicit formula is given in Eq.~\eqref{eq:vNentropy},
is plotted as a function of the number of particles for eight different values 
of the relative interaction strength $\dl/k$. Note that for each value of the 
relative strength there is a peak around the critical value $N_c$.}
\label{graf:thirdbis}
\end{figure}

\begin{figure}[ht] 
\centering
\includegraphics[width=10cm]{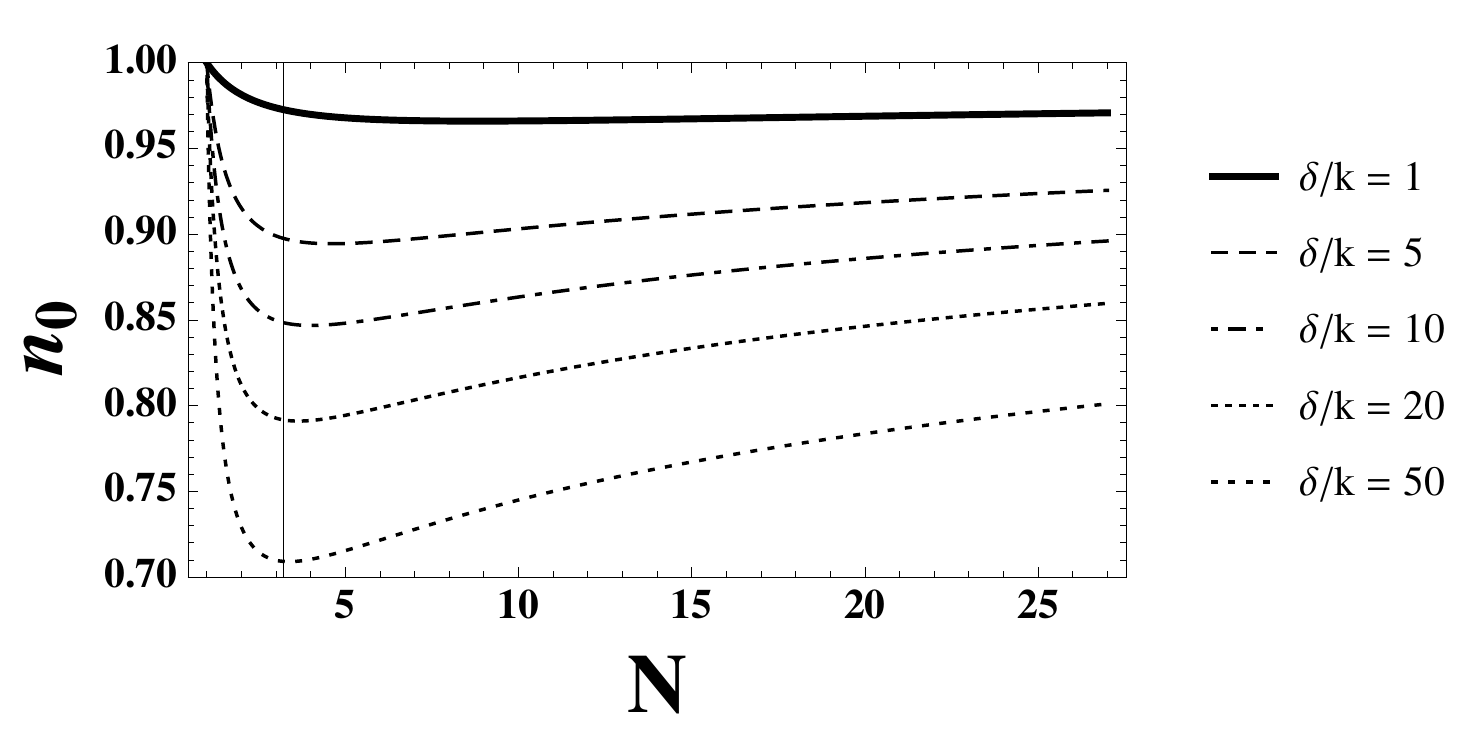} 
\caption{Plot of the first occupation number $n_0 = 1 -t_N$ as a function of the number of 
particles for different \textit{positive} values of $\dl/k$. As it is displayed, $n_0$ takes 
its minimum value around a critical value in the number of particles ($\sim 3.5$).}
\label{graf:occupnum}
\end{figure}

This entanglement measure is numerically examined in Figure \ref{graf:third} in 
terms of the number $N$ of bosons and $\dl/k$. As it is stated in 
\eqref{eq:maximum} the minimum value of this relative strength is $-1/N$. 
There is a critical point around $N_c \sim 3.5$, where the entanglement acquires 
its maximum value. This is more clearly seen 
in Figure \ref{graf:thirdbis}, where the explicit dependence of the von Neuman 
entropy on $N$ is shown. To investigate it in a closer way, we study 
the dependence of the occupation numbers on the number of bosons and the relative 
interaction strength. In particular, we plot the first occupation number separately in 
terms of $N$ and the relative strength. This is done in Figure \ref{graf:occupnum}. 
Therein, we note that below a critical value of number of particles (around 
$N_c \sim 3.5$), for \textit{positive} values of the coupling constant $\dl/k$, 
the first occupation number decreases as the number of particles increases. 
Beyond this value the situation is reversed: the value of the first occupation number 
increases as the number of particles increases.  This implies that $n_0 \sim 1$ in the limit 
when the number of particles tends to infinite, which is a necessary condition for 
the existence of a Bose-Einstein condensation \cite{Pethick}.  

In summary, the critical point around $N \sim 3.5$ where the von Neumann entropy 
is maximum is closely connected with the minimum of the first occupation number 
occurring at such position. It is found 
that above this critical value the spatial entanglement decreases when we add 
up more and more particles to the system, so that the degrowth rate increases 
when the coupling constant is increasing. Moreover, it can be shown that the 
entropy vanishes when $N$ goes to infinite. 

% \S 4
\section{The spinless N-fermion harmonium: linear entropy}
\label{sec:fermions}

In this section we study the spatial entanglement of the N-fermion harmonium 
(i.e., entanglement without taking into account the spin degree of freedom of 
the constituents) using the linear entropy of the one-particle density matrix 
as measure of entanglement of one particle. This 
measure is a first order approximation of the von Neumann entropy of the 
reduced one-body density of the system. In fact, it is a lower bound of this 
entropy. We will consider the case in which the spinless fermions are confined 
in a one-dimensional well, mainly because the antisymmetric fermion case 
requires more elaborate computations. Here we have to consider the 
antisymmetry under exchange of position coordinates $\{r_1,\dots,r_N\}$. 

Taking into account the Hamiltonian \eqref{eq:Mosh-atom} for the N-fermion 
harmonium, one has that the ground-state energy of the system is
\begin{align}
E^f_\gs = \half \om + \sum^{N-1}_{j=1}\mu \big(m +\half\big) = \half \om + \half \mu (N^2-1)
\end{align}
and the corresponding eigenfunction can be expressed as 
\begin{align}
\Psi^f(\xi_1, \dots,\xi_N) =   \frac{1}{\sqrt{N!}} \sum_{J\in S_N} (-)^J J \bigg[
 \phi^\om_{n_N}(\xi_N) \prod^{N-1}_{m=1}\phi^\mu_{n_m}(\xi_m)\bigg],
\label{eq:spinless}
\end{align}
where $n_m \in \{0,\cdots,N-1\}$, so that $n_i \neq n_j$ whenever $i \neq j$.
The symbol $J$ denotes an element of the permutation group $S_N$ of 
$N$~elements (acting on the $r$-coordinates), and $\phi^{\nu}_n$ is the 
single-particle wave function given by an Hermite function of degree $n$ 
with frequency $\nu$. The collective mode $\xi_N$ is symmetric under 
any exchange of the position coordinates. Therefore, one has
$n_N = 0$; otherwise one would have $n_m = 0$ for some $m \neq N$, 
and the wave function would not be totally antisymmetric. This eigenfunction 
can be rewritten as
\begin{align}
\Psi^f_\gs(\xi_1,...,\xi_N) = ({\rm const.}) \sum_{J\in S_N} (-)^J J \bigg[
\prod^{N-1}_{m=1}\ H_{n_m}(\sqrt{\mu} \, \xi_m) \bigg] e^{-\frac{\om}2 \xi_N^2 - 
 \frac{\mu}2 \sum^{N-1}_{m=1}\xi^2_m}, 
 \label{eq:fermion}
\end{align}
where $H_n$ is the Hermite polynomial of degree $n$ and now $n_m \in \{1,\cdots,N-1\}$. 
The exponential power of \eqref{eq:fermion} is the bosonic wave function, 
counterpart of the Wigner function in \eqref{eq:wigner1}. 

As an illustrative case let us consider the 3-harmonium system. To produce an 
antisymmetric wave function there is only one possibility for the degree of the 
polynomials, namely $n_1 = 1$ and $n_2 = 2$. Then the ground-state eigenfunction is
\begin{align}
\Psi^f_\gs(\xi_1,\xi_2,\xi_3) = ({\rm const.}) \sum_{J\in S_3} (-)^J  J\big[
\psi(\xi)\big] 
\, e^{-\frac{\om}2 \xi_3^2 - 
 \frac{\mu}2 (\xi^2_1+\xi_2^2)},
\end{align}
where $\psi(\xi) := H_1(\sqrt{\mu}\,\xi_1) H_2(\sqrt{\mu}\,\xi_2)$
and $\xi := (\xi_1,\xi_2)$.
Note that the six elements of the permutation group $S_3$ are
\begin{align*}
p_1 = \begin{pmatrix}
1 & 2 & 3
\end{pmatrix}, \; \; 
p_2 = \begin{pmatrix}
2 & 1 & 3
\end{pmatrix} , \; \; 
p_3 = \begin{pmatrix}
2 & 3 & 1
\end{pmatrix}, \\
p_4 = \begin{pmatrix}
1 & 3 & 2
\end{pmatrix}, \; \; 
p_5 = \begin{pmatrix}
3 & 1 & 2
\end{pmatrix}, \; \; 
p_6 = \begin{pmatrix}
3 & 2 & 1
\end{pmatrix} .
\end{align*}
The coordinate $\xi$ transforms under the representation of 
the permutation group in the following form:
\begin{align*}
R(p_1) \xi &= \xi, \;\;  R(p_2) \xi = M_1 \xi, \; \;
R(p_3) \xi = M_1 M_2 \xi, \\
R(p_4) \xi &= M_2 \xi, \;\; R(p_5) \xi =  M_1M_3 \xi, \;\;
R(p_6) \xi = M_3 \xi, 
\end{align*} 
where 
\begin{align*}
M_1 = \begin{pmatrix}
-1 & 0 \\
0 & 1 
\end{pmatrix}, \;\; 
M_2 = \begin{pmatrix}
\cos \pi/3 & \sin \pi/3 \\
\sin \pi/3 & -\cos \pi/3
\end{pmatrix} \; \; 
M_3 = \begin{pmatrix}
\cos \pi/3 & -\sin \pi/3 \\
-\sin \pi/3 & -\cos \pi/3
\end{pmatrix} .
\end{align*}
Note that $\det M_i = -1$ and $M^t M = 1$. 
Any matrix 
$$
M_\phi = \twobytwo{\cos\phi}{\sin\phi}{\sin\phi}{-\cos\phi}
$$
is a reflection in the axis $\theta = \phi/2$. In our case, this means
$\theta = \pi/2, \pi/6, -\pi/6$; i.e., three reflection axes at 
60-degree angles to each other. The rotations 
$R(p_3)$ and $R(p_5)$ are products of two reflections each.
Thus, the eigenfunction transforms as follows: 
\begin{align}
\Psi^f_\gs(\xi_1,\xi_2,\xi_3) &=  ({\rm const.}) \, 
e^{-\frac{\om}2 \xi_3^2 - \frac{\mu}2 (\xi^2_1+\xi_2^2)} \, \sum^6_{i=1} (-)^{i+1} \psi( R(p_i) \xi) 
\nonumber \\
&= ({\rm const.})\, e^{-\frac{\om-\mu}2 \xi_3^2 - \frac{\mu}2 (r_1^2 +r_2^2+r_3^2)} \, 
\sum^6_{i=1} (-)^{i+1} \psi( R(p_i) \xi) \nonumber \\
&= ({\rm const.})\, e^{-\frac{\om-\mu}2 \xi_3^2 - \frac{\mu}2 (r_1^2 +r_2^2+r_3^2)} \, 
(r_1-r_2)(r_1-r_3)(r_2-r_3).
\end{align}
Notice that the last equality has been possible because
\begin{align}
\frac1{12\sqrt{2}}\frac{1}{\mu^{3/2}}\sum^6_{i=1} (-)^{i+1} \psi(\sqrt{\mu}\, R(p_i) \xi) 
=  - \frac1{\sqrt{2}} (\xi_1^3 - 3 \xi_1 \xi_2^2) = (r_1-r_2)(r_1-r_3)(r_2-r_3) 
\end{align}
is a Vandermonde determinant. This 3-fermion result can be extended to the 
N-fermion system. In fact, general results of the theory of antisymmetric functions 
ensure that the wave function of a system of spinless fermions is equal to the 
wave function of the corresponding bosonic system multiplied by the $N$-variable
Vandermonde determinant (see Section 3.1 of \cite{Procesi}). 
So, the ground-state wave function of the spinless N-fermion harmonium 
described by the Hamiltonian 
\eqref{eq:Mosh-atom} has the form:
\begin{align}
\Psi^f(r_1,\cdots,r_N) =  ({\rm const.})
\prod_{ i\,<\,j} (r_i - r_j) \, \exp\bigg[- \frac{\om - \mu}{2} \,
\xi^2_N - \frac{\mu}{2} \sum^N_{i=1} r_i^2\bigg],
\label{eq:fermions}
\end{align}
where the product-like symbol on the right hand of this expression denotes 
the Vandermonde determinant:
\begin{align}
\V_{(r_1, \cdots, r_N)} := \begin{vmatrix}
1               & \cdots & 1 \\
r_1           & \cdots & r_N \\
\vdots       & \ddots & \vdots \\
r_1^{N-1} & \cdots & r_N^{N-1} 
\end{vmatrix}
 = \prod_{1\leq i\,<\,j\leq N} (r_i - r_j) .
\end{align}
Concrete calculations \cite{LosChinos, Johnson} have borne out this 
fact several times. On the other hand, it has been pointed out by 
several authors that this function is actually a generalization of Laughlin's 
wave function for the fractional quantum Hall effect\cite{Hall}.  This is not 
surprising since in the Hall effect the magnetic field can be understood as a
harmonic potential acting on the electrons. Let us also point out that the Wigner 
function corresponding to the Gaussian one on the right-hand side 
of~\eqref{eq:fermions} is nothing but the Wigner N-body quasidensity 
given by \eqref{eq:wigner1}.

Let us now calculate the one-body density of the one-dimensional spinless 
N-fermion harmonium, which is defined by 
\begin{align}
\rho_1(r;r') = 
\int dr_2 \cdots dr_N \; \Psi^f(r,r_2,\cdots, r_N) \Psi^f(r',r_2,\cdots, r_N) .
\label{eq:rho1}
\end{align}

For this purpose it is convenient to rewrite the eigenfunction \eqref{eq:fermions} as:
\begin{align}
\Psi^f(r_1,\cdots, r_N) &=
({\rm const.}) \, \V_{(r_1, \cdots, r_N)}
 \, e^{-a(r^2_1+\cdots +r_N^2) + b_N(r_1+\cdots+r_N)^2},
\end{align}
with $a =\frac{\mu}2$ and $b_N = \frac{\mu-\om}{2N}$. Moreover, we will 
use the Hubbard-Stratonovich transformation 
\begin{align}
e^{\al\zeta^2} = \sqrt{\frac{\al}{\pi}} \int^{\infty}_{-\infty} dz\, e^{-\al z^2+2\al z\zeta},
\word{with} \al\in \C \word{and} {\rm Re}(\al) > 0,
\end{align}
where $\zeta = r_2 + \cdots + r_N$ and $\al = 2b_N$.  Then the one-body density reads:
\begin{align}
\rho_1(r;r')  &= 
 k_N \, e^{-(a-b_N)(r^2+r'^2)} \int dz \, e^{-2b_N z^2} 
e^{(N-1)\frac{b_N^2}{2a}(r+r'+2z)^2}  \nonumber  \\
&\qquad \x \int dz_2 \cdots dz_N 
  \V_{(z_r,z_2, \cdots, z_N)} \, \V_{(z'_{r},z_2, \cdots, z_N)} \,
\prod_{j=2}^N e^{- z_j^2},
\end{align}
where $k_N$ denotes the normalization constant (to be calculated later), 
and $z_j := \sqrt{2a}\big[r_j - \frac{b_N}{2a}(r+r'+2z)\big]$ is a convenient 
change of coordinates. A key observation \cite{Schilling} 
is that the last expression on the right-hand side 
is the multiplication of two Slater determinants and 
therefore, the computation of the reduced density matrix can be done 
using well-known methods in quantum chemistry. Indeed, with the notation
\begin{align}
&\beta_N := \sqrt{\frac{\mu-\om}{(N-1)\om+\mu}}, \qquad
q_{(r,r')} := \sqrt{\mu}\big[r - \half \beta_N^2(r+r')\big] \nonumber
\\ 
& \qquad \word{and} c_N := \frac{(\mu-\om)^2}{(N-1)\om+\mu}\frac{N-1}{2N},
\label{eq:defintions}
\end{align}
one  can express the one-body density as
\begin{align}
\label{eq:e7}
\rho_1(r;r') =
 k_N \, e^{-a_N (r^2+r'^2) +2c_N rr'}   \int du \, e^{-u^2} 
\sum^{N-1}_{j=0} \frac{1}{2^j j!}H_j[q_{(r,r')} -\beta_N u] H_j[q_{(r',r)} - \beta_N u] ,
\end{align}
with $a_N := a-b_N-c_N$. This expression can be calculated analytically 
with the help of the following general formula derived in Appendix \ref{apendicea}: 
\begin{align}
&\int_{-\infty}^{\infty} e^{-u^{2}}  H_{k}(q_{(r,r')}-\beta_N u)H_{k}(q_{(r',r)}- \beta_N u)\, du 
\nonumber \\
&  = \sum_{\substack{n_{1},n_{2}=0 \\ n_1+n_2 \, {\rm even}}}^{k} 
(2\beta_N)^{n_{1}+n_{2}}  {k \choose n_{1}} {k \choose n_{2}} 
H_{k-n_{1}}(-q_{(r,r')}) H_{k-n_{2}}(-q_{(r',r)}) \, \Gamma \left(\frac{n_{1}+n_{2}+1}{2}\right). 
\end{align}
Then the integral at the right-hand side of \eqref{eq:e7} can be expressed 
as the following power sum:
\begin{align}
\int du \, e^{-u^2} 
\sum^{N-1}_{j=0} \frac{1}{2^j j!}H_j[q_{(r,r')} -\beta_N u] H_j[q_{(r',r)} - \beta_N u]  = 
 \sqrt{\pi} \sum^{N-1}_{t=0} \sum^{2t}_{s=0}
  c^N_{t,s} \, r^{2t-s}r'^s ,
\end{align}
where the expansion coefficients $c^N_{r,s}$ depend on the frequencies $\om$, $\mu$ and the 
number of particles. When $r + s$ is an odd number or bigger than $2(N-1)$,
$c^N_{r,s} = 0$.
By symmetry it is clear that $c^N_{t,s} = c^N_{s,t}$.
In the one fermion case there is only one coefficient
$c^1_{0,0} = 1$. For $N$ fermions, the rank of the quadratic form of the $c_{r,s}$ is 
$2N-1$. In particular, when $N = 2$, we have the expansion coefficients
\begin{align}
c^2_{0,0} = \frac{2\mu}{\om + \mu}, \quad
c^2_{1,1} = \frac{
\mu (\mu^2 + 2 \mu \om +5\om^2)}{(\om+\mu)^2} \word{and}
c^2_{2,0} = c^2_{0,2} = \frac{\mu(\om-\mu)(3\om+\mu)}{2(\om+\mu)^2} .
\end{align}
For completeness, let us mention that the reader will have no difficulty in 
verifying that $\rho_1(r,r)$ ---the diagonal 
of \eqref{eq:rho1}--- coincides (except for the difference in the 
normalization conventions) with the one obtained in \cite[Sect.~V]{Laetitia} 
 by the Wigner function method. Moreover, for $N = 3$ the non-vanishing expansion coefficients are:
\begin{align}
c^3_{0,0} &= \frac32 \bigg[1+ \frac{(\om -\mu)^2}{(2\om + \mu)^2}\bigg] , \quad
c^3_{1,1} = \frac{\mu(-15\om^3+51 \mu\om^2+15\mu^2\om+3\mu^3)}{(2\om+\mu)^3}, 
\nonumber \\
& c^3_{2,2} = \frac{\mu^2(363\om^4 + 168 \mu\om^3+90\mu^2\om^2+24\mu^3\om+3\mu^4)}
{4(2\om+\mu)^4}, \nonumber \\
c^3_{2,0} = c^3_{0,2} &= -\frac{\mu(39\om^3-3 \mu\om^2+
15\mu^2\om+3\mu^3)}{2(2\om+\mu)^3}, \quad
c^3_{4,0} = c^3_{0,4} =  \frac{\mu^2(5\om+\mu)^2(\om-\mu)^2}{8(2\om+\mu)^4} \nonumber \\
& \word{and} c^3_{3,1} = c^3_{1,3} =  \frac{\mu^2(\om-\mu)(65\om^3+33
\mu\om^2 + 9 \mu^2\om + \mu^3)}{2(2\om+\mu)^4} .
\end{align}
Except for the difference in the normalization and notation conventions, 
these coefficients coincide with the ones found in \cite{Schilling}.

Keeping all this in mind and determining the normalization constant $k_N$ 
by imposing that $\int \rho_1(r;r) \, dr = 1$, one finally has the expression 
\begin{align}
\rho_1(r;r') =
\frac{1}{\sqrt{\pi}}\frac1{\sqrt{N}} \sqrt{\frac{\om\mu}{(N-1)\om + \mu}} \, e^{-a_N (r^2+r'^2) +
2c_N rr'}    \sum^{N-1}_{t=0} \sum^{2r}_{s=0}
  c^N_{t,s} \, r^{2t-s}r'^s  
\end{align}
for the one-body density of the one-dimensional spinless N-fermion harmonium.

Let us now quantify the entanglement of the system by means of the linear 
entropy associated to the one-body density \cite{Manzano}:
\begin{align}
S_L = 1 -  N \Tr[\rho_1^2],
\label{eq:linearentropy}
\end{align}
where $\Pi_N = N \Tr[\rho_1^2]$ is the \textit{purity} of the system. 
This entanglement measure, which is a non-negative quantity that vanishes 
if and only if the state has Slater rank 1 and it is therefore separable, has been 
recently used in various two-fermion systems \cite{YAN, MANZA} as well as for
 various helium-like systems \cite{Dehesa, HOFER, LIN} in both ground and excited 
 states. Let us also point out that the linear entropy is a linearization of the von Neumann 
 entropy, and gives a lower bound for this logarithmic entropy. Moreover, in our 
 systems it turns out that
\begin{align}
 \Tr[\rho_1^2] = \int \rho_1(r;r')\rho_1(r';r) \, dr' \, dr  &= 
 k^2_N 
\sum^{N-1}_{t,e=0} \sum^{2t}_{s=0} \sum^{2e}_{f=0}
 c^N_{t,s} c^N_{e,f} \, \mathcal{I}_{(2t-s+f,2e-f+s)},
 \label{eq:otra}
\end{align}
where, as shown in Appendix \ref{gaussianas}, 
\begin{align}
\mathcal{I}_{(n,m)} &= \mathcal{I}_{(m,n)} \nonumber \\
&= \int^{\infty}_{-\infty} r^n r'^m 
e^{-2a_N (r^2+r'^2) +4C_N rr'} \, dx \, dy \nonumber \\
&=
2 \sum^n_{i=0}\sum^m_{j=0}
\binom{n}{i}\binom{m}{j} (-1)^j  
\la_1^{-(n+m-i-j+1)/2}\la_2^{-(i+j+1)/2} \nonumber \\ & \quad \x
\Ga\Big(\frac{n+m-i-j+1}{2}\Big)\Ga\Big(\frac{n+m+1}{2}\Big)
\frac14 [1+(-1)^{n+m-i-j}] [1+(-1)^{i+j}],
\label{eq:purity}
\end{align}
with $\la_1 = 4(a_N-c_N)$ and $\la_2 = 4(a_N+c_N)$.

\begin{figure}[ht] 
\centering
\begin{subfigure}[b]{.48\textwidth}
 \centering
 \includegraphics[width=7.8cm]{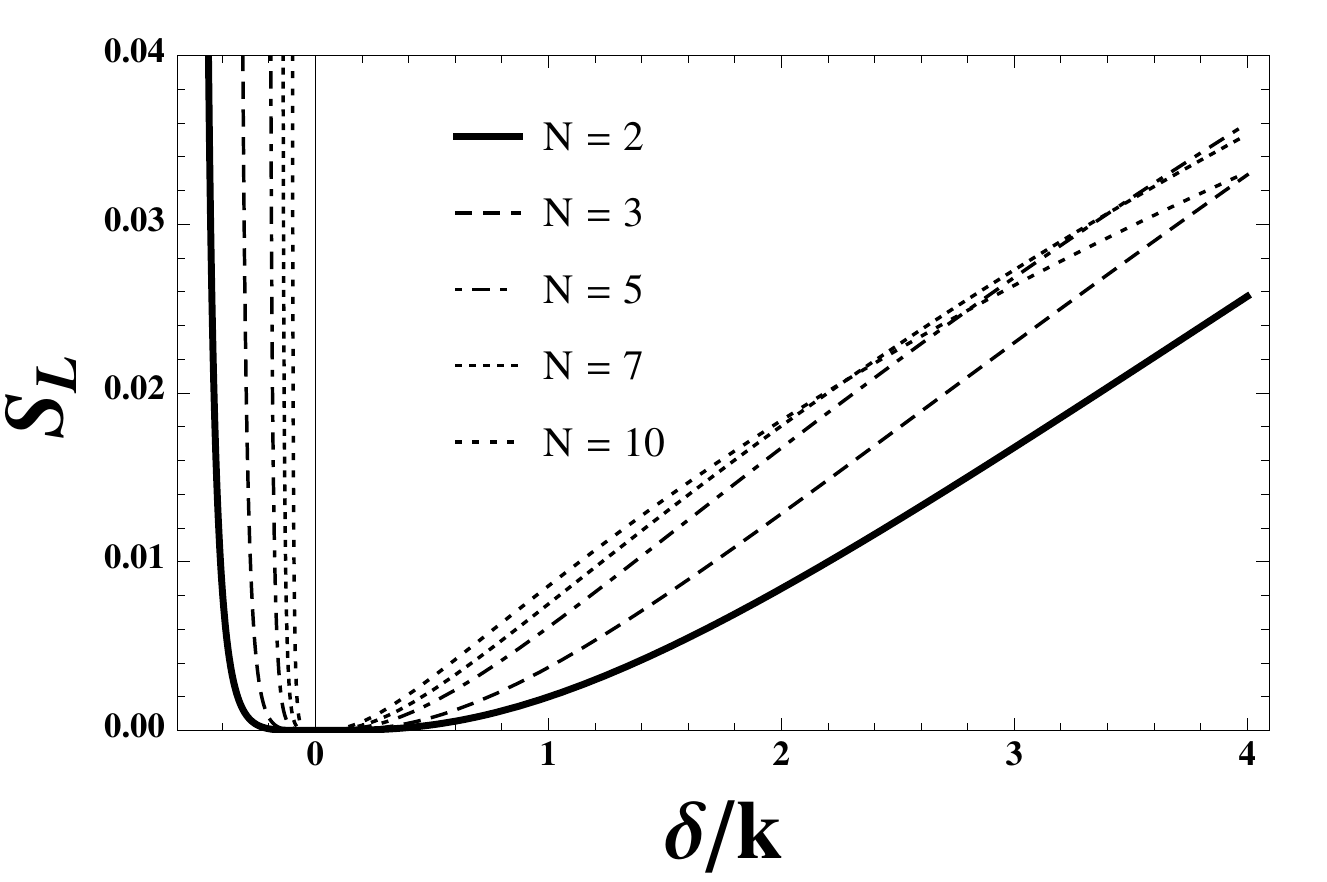} 
%\caption{} 
\end{subfigure}
\hfill
\begin{subfigure}[b]{.48\textwidth}
\centering
\includegraphics[width=7.8cm]{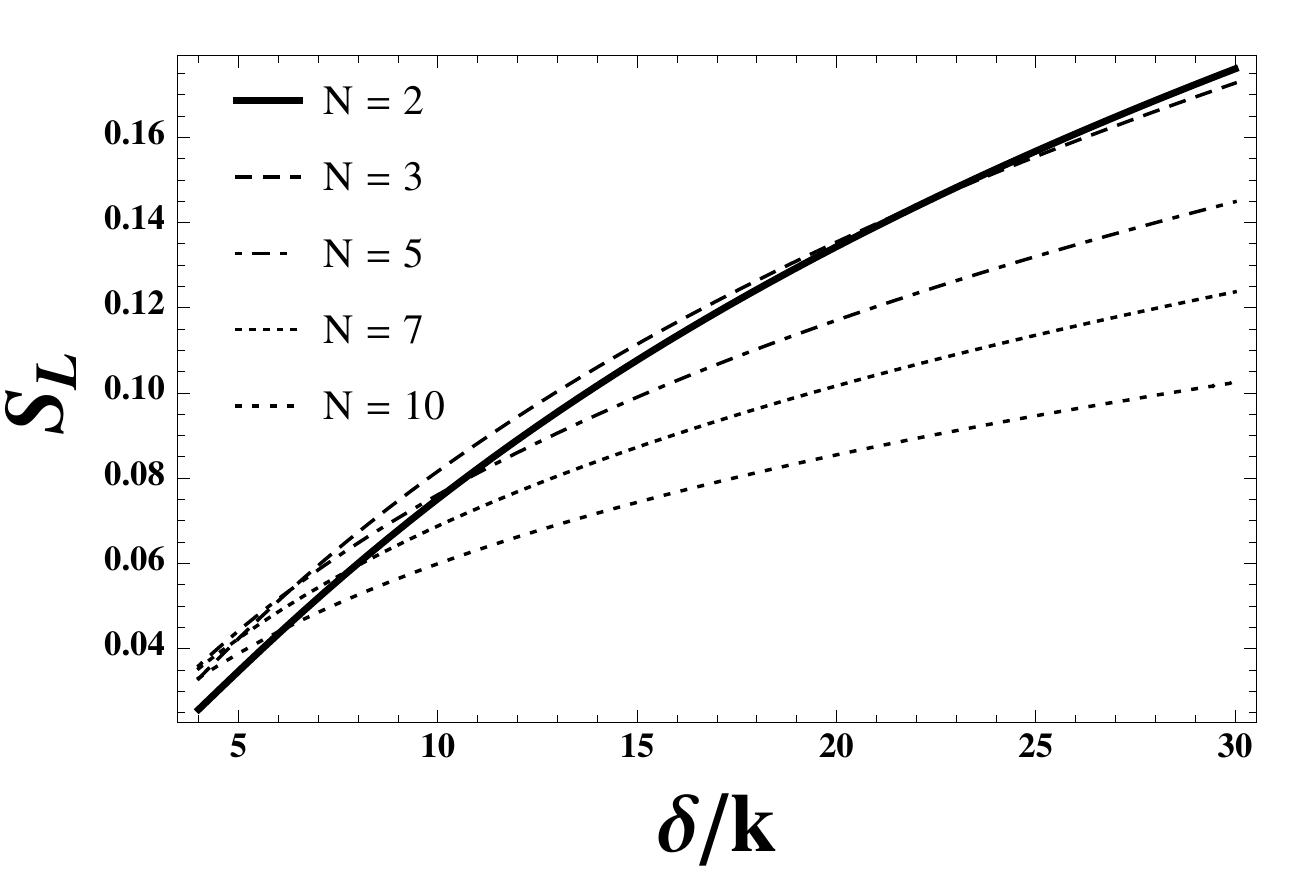}  
%\caption{}
\end{subfigure}
\hfill \caption{Linear entropy of the one-particle density matrix 
 of the N-fermion harmonium as a function of the coupling constant for five different values 
 of $N$. Note the change of behavior when the strength grows.}
\label{graf:five}
\end{figure}

Using expressions \eqref{eq:linearentropy},  \eqref{eq:otra} and \eqref{eq:purity}, 
we have investigated the dependence 
of the linear entropy of the system on the relative interaction strength $\dl/k$, 
as well as on the energy. The results are given in Figure 
\ref{graf:five}. A first basic observation is that for fixed $N$ the 
entanglement decreases (increases) when the relative interaction strength 
is increasing in the negative (positive) region. Moreover, for negative values of 
the coupling constant, the spatial entanglement of the N-fermion harmonium grows 
when $N$ is increasing. For small positive values of the coupling 
constant  (i.e., when the fermions attract each other); herein we find that for very 
small values of $\dl/k$ the entanglement grows again with increasing $N$. In general,
we observe that for positive values of the strength we 
have various regimes of dependence on $N$. In the strong-coupling regime, the 
linear entropy grows as the number of particles is increasing since the purity 
decreases with increasing number of particles. For instance, for $\dl/k =10$ the 
purity is 0.92 for $N=2$ and 0.94 for $N = 10$, whereas for $\dl/k =1$ is 0.998 
and 0.992 respectively.

\begin{figure}[ht] 
\centering
\includegraphics[width=9cm]{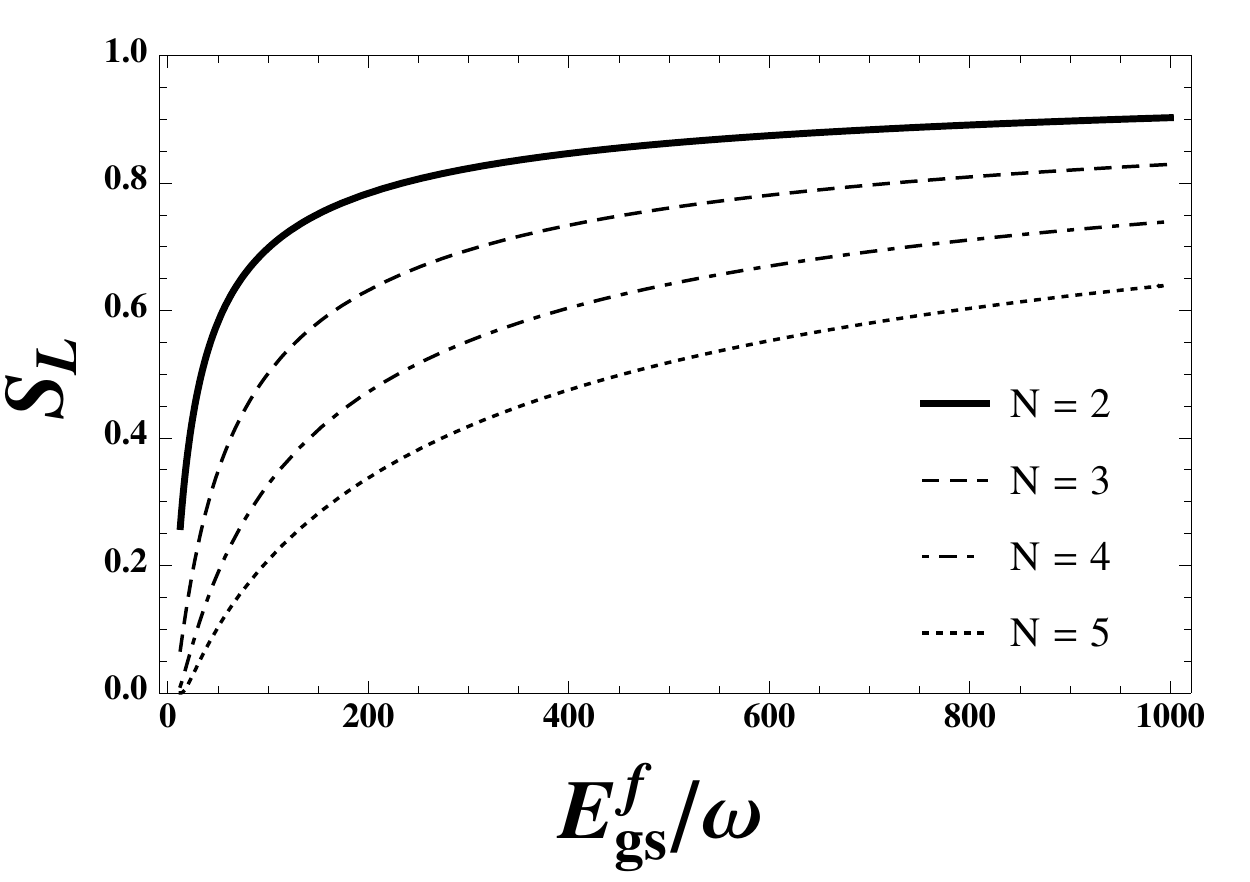}  
\hfill \caption{Linear entropy $S_L$ of the N-fermion harmonium versus the ground-state 
energy $E^f_\gs/\om$ for different values of $N$. The relation between the relative strength
$\dl/k$ and the energy is stated in \eqref{eq:relativestrenght}. The entanglement grows when the energy increases.}
\label{graf:six}
\end{figure}

Figure \ref{graf:six} displays the dependence of the linear entropy of the 
system on the ground state energy $E^f_\gs/\om$ for various values of the 
number of fermions $N$. We should keep in mind that the relation between 
the relative interaction strength and the energy is given by
\begin{align}
\dl/k = \bigg[\frac{2 E^f_\gs/\om-1}{N^2-1} \bigg]^2-1.
\label{eq:relativestrenght}
\end{align}
We find that the entanglement grows when the energy increases. For $N = 2$ 
this behavior has been recently observed not only in the Moshinsky or 2-fermion 
harmonium atom but also in other 2-fermion systems \cite{MANZA, Dehesa, 
Benenti, HOFER, LIN}. The novelty is that this behavior appears to be true 
also for heavier N-fermion harmonium atoms.

It is worth noting that our results can be 
considered as a generalization of previous results for the entanglement 
entropy of the fractional quantum Hall effect based in an exact matrix-product
representation of the Laughlin wave function \cite{Laugh1, Laugh2, Laugh3}. 
This well-known wave function, formulated by R. Laughlin to explain the fractional 
quantum Hall effect, consists of a lone Slater determinant and 
reads as in \eqref{eq:fermions} except that there is no coupling between the
electrons, that is, $\om = \mu$ or $\dl = 0$, and the domain is the complex space. 

% \S 4
\section{Total entanglement of the spinned N-fermion harmonium} 
\label{sec:spin}

So far, we have considered only spinless particles. This means that in 
the two previous sections we have determined the spatial entanglement 
of the N-boson and N-fermion harmonia. Nevertheless, our conclusions 
for the bosonic system also apply to the total entanglement from a 
qualitative point of view, because the spin part of the ground-state 
eigenfunction fully factorizes. For fermionic systems this is no longer true, 
at least when they are isolated. In fact, 
for the ground state of two fermions is the product of a two-boson state and
an antisymmetric spin state. For a system of fermions with spin in presence
of a strong magnetic field, the spin part of the eigenfunction factorizes and 
so the spatial part must be antisymmetric  \cite{Schilling}. In general, however, 
for the isolated fermionic system the spatial and spin parts of the ground-state 
wave function are not separable, and antisymmetry must take into consideration 
both spin and position coordinates.

Let us now calculate the total entanglement of the N-fermion harmonium (i.e., 
the entanglement of both spatial and spin degrees of freedom) using the linear 
entropy as measure of entanglement. For simplicity, we consider one-dimensional 
models with an even number (say $2N$) of spinned electrons from now on. 
The total spin of the system is zero, and all the spatial orbitals are doubly 
occupied (restricted configuration).
This will eventually allow for an easy generalization to closed-shell three-dimensional 
systems. We define the set of spin natural orbitals in the following way:
\begin{equation*}
\vf_j (x) = \frac{}{} \left\{
\begin{array}{rl}
\phi_{\frac{j-1}{2}}(r) \, \up, & \text{if } j  \; \text{is odd}, \\
\phi_{\frac{j  }{2}}(r) \, \dn, & \text{if } j \; \text{is even}.
\end{array} \right .
\end{equation*}
The symbols $\up$ and $\dn$ mean spin up and spin down, respectively, and we denote $x := (r, \vs)$, $\vs$ being the spin coordinate. 

Let us begin with the consideration of the simpler non-interacting case; that is, 
when $\dl = 0$, and hence $\mu = \om$.  The corresponding ground-state eigenfunction 
is then given by the expression
\begin{align}
\Psi^f(x_1,\ldots,x_N) &= \frac{1}{\sqrt{2N!}}
\sum_{J\in S_{2N}} (-)^J 
\Bigg[\prod^{2N}_{i=1} \vf_{i-1}(x_{J(i)})\Bigg] \nonumber \\
&=   \frac{1}{\sqrt{2N!}} \sum_{J\in S_{2N}} (-)^J
\, J
\Bigg[\prod^{N}_{i=1} \up_{2i-1} \dn_{2i} \phi^\om_{i-1}(r_{2i-1}) \, 
\phi^\om_{i-1}(r_{2i})\Bigg] .
\label{eq:semifinal}
\end{align}
In the non-interacting case, the basis of the one-particle Hilbert space 
is the set of Hermite functions whose degree is less than $N$. Moreover,
%In the permutation
%group $S_{2N}$ there are subgroups of permutations whose effect in the spin 
%part of the expression \eqref{eq:semifinal} coincide. 
let us define 
\begin{align*}
H &= \{j\in S_{2N} : j \; {\rm fixes} \; 1, \ldots, N\} \simeq S_N \\
K &= \{j\in S_{2N}: j \; {\rm fixes} \; N+1,\dots, 2N \} \simeq S_N
\end{align*}
and $\tilde{S}_{N} = H\x K$ the direct product of these two subgroups. 
The set $S'_{2N} = S_{2N}/\tilde{S}_{N}$ is the set of right cosets in 
$S_{2N}$, giving the following equivalence relation: $J \sim J'$ if and only if 
there exists $(j,j') \in \tilde{S}_{N}$ such that $J' = (j,j') J$. 

Therefore we can reorganize the expression \eqref{eq:semifinal} in the following form:
\begin{align}
\Psi^f(x_1,\ldots,x_N) &=  \frac{1}{\sqrt{2N!}}
\sum_{} (-)^J
\prod^{N}_{i=1} \up_{J(i)} \prod^{2N}_{i=N+1} 
\dn_{J(i)} \nonumber \\ &\x
\begin{vmatrix}
\phi^\om_0(r_{J(1)})   & \cdots & \phi^\om_0(r_{J(N)})  \\
\vdots               & \ddots & \vdots \\
\phi^\om_N(r_{J(1)})  & \cdots & \phi^\om_N(r_{J(N)})  
\end{vmatrix}  \begin{vmatrix}
\phi^\om_0(r_{J(N+1)})   & \cdots & \phi^\om_0(r_{J(2N)})  \\
\vdots               & \ddots & \vdots \\
\phi^\om_N(r_{J(N+1)})  & \cdots & \phi^\om_N(r_{J(2N)})  ,
\end{vmatrix} ,
\label{eq:twoslater}
\end{align}
where the sum runs over a representative of each coset. 
The choice of the representative $J\in [J] \in S'_{2N}$ is immaterial.
The number of summands is 
$$
\begin{vmatrix} S_{2N}/\tilde{S}_N\end{vmatrix} = \frac{|S_{2N}|}{|\tilde{S}_{N}|} =
\frac{(2N)!}{(N!)^2} = \binom{2N}{N}. 
$$
 For instance, for the four-harmonium ($N = 4$), the eigenfunction can be 
 written in the alternative form:
 \begin{align*}
\Psi^f(x_1,x_2,x_3,x_4) &=   \frac{\om^2}{\pi}
\frac{1}{\sqrt{3!}} e^{-\frac{\om}2 (r^2_1+r^2_2+r^2_3+r^2_4)}
[
(\up_1 \up_2 \dn_3 \dn_4 + \dn_1 \dn_2 \up_3 \up_4)\V_{(1,2)}\V_{(3,4)} \\&-
(\up_1 \up_3 \dn_2 \dn_4 + \dn_1 \dn_3 \up_2 \up_4)\V_{(1,3)}\V_{(2,4)} 
+(\up_1 \up_4 \dn_2 \dn_3 + \dn_1 \dn_4 \up_2 \up_3)\V_{(1,4)}\V_{(2,3)}].
\end{align*}
The one-body reduced density matrix corresponding to the wave function \eqref{eq:twoslater}
is diagonal in spin space and is given by the expression 
\begin{align}
\rho_1(x,x') =
\begin{pmatrix}\rho^{\up\up}_1(r,r') & 0  \\
0 & \rho^{\dn\dn}_1(r,r')
\end{pmatrix} .
\end{align}
Moreover, it is clear that   
\begin{align}
\rho^{\up\up}_1(r,r') &= \rho^{\dn\dn}_1(r,r') 
=  \int  dx_2 \cdots dx_{2N} \, 
\Psi^f(r,\vs_1,x_2,\ldots,x_N) \Psi^f(r',\vs_1,x_2,\ldots,x_N) |_{\vs_1=\up} \nonumber 
\\
&= \frac{1}{2N} \frac1{(N-1)!}\int  dr_2 \cdots dr_{N}
\begin{vmatrix}
\vf_0(r)  & \cdots & \vf_0(r_N)  \\
\vdots        & \ddots & \vdots \\
\vf_N(r)  & \cdots & \vf_N(r_N)  
\end{vmatrix}   \begin{vmatrix}
\vf_0(r')  & \cdots & \vf_0(r_N)  \\
\vdots        & \ddots & \vdots \\
\vf_N(r')  & \cdots & \vf_N(r_N)  
\end{vmatrix}  \nonumber \\
&= \frac{1}{2N} \sum^{N-1}_{i=0} \vf_i(r)\vf_i(r').
\end{align}
The occupation numbers appear twice, what is a well-known result 
for atomic and molecular scientists.

Let us now consider the general interacting case, for which $\dl \neq 0$, and 
hence $\mu \neq \om$. The ground-state eigenfunction for the Hamiltonian 
\eqref{eq:Mosh-atom} is similar to \eqref{eq:spinless}, except that each spatial 
orbital is doubly occupied. For the same 
reasons as in the spinless case, the collective mode $\xi_{2N}$ occupies the Hermite 
function of degree zero. The other coordinates $\{\xi_m\}^{2N-1}_{m=1}$ occupy the other 
spatial orbitals in such a way that the total wave function is totally antisymmetric under
 interchange the coordinates $\{x_m\}^{2N}_{m=1}$. The ground-state eigenfunction 
 schematically reads
\begin{align}
\Psi^f(x_1,...,x_{2N}) =  
\frac{1}{\sqrt{2N!}}
\sum_{J\in S_{2N}} (-)^J \bigg[
\prod^{N}_{i=1} \up_{J(i)} 
 \prod^{2N}_{i=N+1} \dn_{J(i)}  \bigg]
 J  \bigg[\phi^\om_{0}(\xi_{2N})\phi^\mu_{0}(\.)
 \prod^{N}_{m=1}\phi^\mu_{m}(\.)\phi^\mu_{m}(\.)\bigg].
\end{align}

For instance, in the particular case $N$ = 4 the eigenfunction has the form
\begin{align*}
\Psi^f(x_1,x_2,x_3,x_4) &=  
\frac{1}{\sqrt{4!}}
\sum_{J\in S_4} (-)^J \up_{J(1)} \up_{J(2)} \dn_{J(3)}  \dn_{J(4)} 
 J  \big[\phi^\om_{0}(\xi_4) \phi^\mu_{1}(\xi_1)
 \phi^\mu_{0}(\xi_3) \phi^\mu_{1}(\xi_2)\big] \\
 &= ({\rm const.})
\sum_{J\in S_4} (-)^J \up_{J(1)} \up_{J(2)} \dn_{J(3)}  \dn_{J(4)} 
 J  \big[\xi_1\xi_2\big] e^{-\frac\om2 \xi_4 - \frac\mu2(\xi_1^2 + \xi_2^2+\xi_3^2)} \\
 & = ({\rm const.})
 \sum_{S'_4} (-)^J \up_{J(1)} \up_{J(2)} \dn_{J(3)}  \dn_{J(4)} 
\V_{(J(1),J(2))} \V_{(J(3),J(4))} e^{-\frac\om2 \xi_4 - \frac\mu2(\xi_1^2 + \xi_2^2+\xi_3^2)}.
\end{align*}
Using again that an antisymmetric polynomial is equal to a symmetric
polynomial multiplied by a Vandermonde determinant \cite{Procesi}, we can 
write the ground-state eigenfunction of the general interacting spinned N-fermion system as:
\begin{align}
\Psi^f(x_1,\ldots,x_{2N}) =&  \, ({\rm const.})\,
e^{- \frac{\om - \mu}{2} \xi^2_{2N} - \frac{\mu}{2} \sum^{2N}_{i=1} r_i^2} \nonumber
\\ &\x\sum_{S'_{2N}} (-)^{J} 
\Bigg[\prod^{N}_{i=1} \up_{J(i)} \prod^{2N}_{i=N+1} \dn_{J(i)} \Bigg] 
\V_{(J(1),\ldots,J(N))}\V_{(J(N+1),\ldots,J(2N))}.
\end{align}
and the corresponding energy is 
$E^f_\gs  = \half (\om+\mu) + \mu (N^2-1)$. Let us 
define $\eta = r + r_2 + \cdots + r_N$ and  $\eta' = r' + r_2+\cdots + r_N$.
To compute the one-density we use twice the Hubbard-Stratonovich identity.
First to compute $N$ integrals with $\zeta'= r_{N+1} + \cdots + r_{2N}$,
and second to compute $N-1$ integrals  with $\zeta= r_2 + \cdots + r_N$.
Each diagonal element of the one-body density matrix reads:
\begin{align}
&\rho^{\up\up}_1(r,r') = \rho^{\dn\dn}_1(r,r') \nonumber \\
&=  
 ({\rm const.}) \, e^{-a(r^2+r'^2)} \int dr_2 \cdots dr_N\,
e^{- 2a(r^2_2 + \cdots + r^2_N)+b_{2N}(\eta^2+\eta'^2)}  
\int dz \, e^{-2b_{2N} z^2} 
e^{N\frac{b_{2N}^2}{2a}(\eta+\eta'+2z)^2} \nonumber \\
&\qquad \x \V_{(r,,2,...,N)}\, \V_{(r',2,\ldots,N)} \int dz_{N+1} \cdots dz_{2N} 
  \, \V_{(z_{N+1},\ldots, z_{2N})}^2 \,
\prod_{j=N+1}^{2N} e^{- z_j^2},
\end{align}
where $z_j := \sqrt{2a}\big[r_j - \frac{b_{2N}}{2a}(\eta+\eta'+2z)\big]$. 
Once again, the expression on the right is the total integral of the 
product of two Slater determinants,
that is, 
$$
\int dz_{N+1} \cdots dz_{2N}  \V_{(z_{N+1},\ldots, z_{2N})}^2 \prod_{j=N+1}^{2N} e^{- z_j^2}
$$
is a real constant. Therefore
\begin{align}
&\rho^{\up\up}_1(r,r') = \rho^{\dn\dn}_1(r,r') \nonumber \\
&= ({\rm const.}) \, e^{-a(r^2 + r'^2)} \int dr_2 \cdots dr_N \,
e^{-2a(r_2^2 + \cdots+ r_N^2) + b_{2N}(\eta^2+\eta'^2)+d_{N} (\eta+\eta')^2}  \,
 \V_{r,,2,...,N}\, \V_{r',2,\ldots,N}  \nonumber \\
% &= ({\rm const.}) \, e^{-a(r^2 + r'^2) + b_{2N}(r^2+r'^2)+d_N(r+r')^2}
% \int dr_2 \cdots dr_N \,
%e^{-2a(r_2^2 + \cdots+ r_N^2) + 2(b_{2N}+ 2d_N) [\zeta'^2+\zeta'(r+r')]}  \\
%& \qquad \x
% \V_{r,,2,...,N}\, \V_{r',2,\ldots,N} \\
 &= ({\rm const.}) \, e^{-a(r^2 + r'^2) + b_{2N}(r^2+r'^2)+d_N(r+r')^2}
 \int dz \, e^{-2 g_N z^2} \, e^{(N-1)\frac{g_N^2}{2a}(r+r'+2z)^2}  \nonumber
 \\
 &\qquad \x \int dz_2 \cdots dz_N \,
 \V_{z_r,,z_2,...,z_N}\, \V_{z'_{r},z_2,\ldots,z_N}
\prod^N_{j=2} e^{-z'^2_j},
\end{align}
where $z'_j := \sqrt{2a}\big[r_j - \frac{g_N}{2a}(r+r'+2z)\big]$,
 $d_N = \frac1{8N} \frac{(\mu-\om)^2}{\mu+\om}$ and $g_N = \frac{\mu-\om}{2N}
 \frac{\mu}{\mu+\om}$. As in the spinless case, let us define
\begin{align}
&\tilde{q}_{(r,r')} = \sqrt{\mu}\big[r - \half \beta_{2N}^2 (r+r')\big] \word{and}
\tilde{c}_{2N} = \frac{(\mu-\om)^2}{(2N-1)\om+\mu}\frac{2N-1}{8N}
\end{align}
with $\bt_N$ as defined in \eqref{eq:defintions}. Then
\begin{align}
&\rho^{\up\up}_1(r,r') = \rho^{\dn\dn}_1(r,r') = 
\frac{1}{2\pi}\frac1{\sqrt{N}} \sqrt{\frac{2\om\mu}{(2N-1)\om + \mu}} \nonumber
\\
&\qquad \qquad \x e^{-a'_N(r^2 + r'^2) + 2 \tilde{c}_{2N} rr' }  \int du \, e^{-u^2} 
\sum^{N-1}_{j=0} \frac{1}{2^j j!}H_j[\tilde{q}_{(r,r')} -\beta_{2N} u] H_j[\tilde{q}_{(r',r)} - \beta_{2N} u] ,
\label{eq:expre}
\end{align}
with $a'_N := a-b_{2N}- \tilde{c}_{2N}$. 
 
 \begin{figure}[ht] 
 \centering
 \includegraphics[width=9cm]{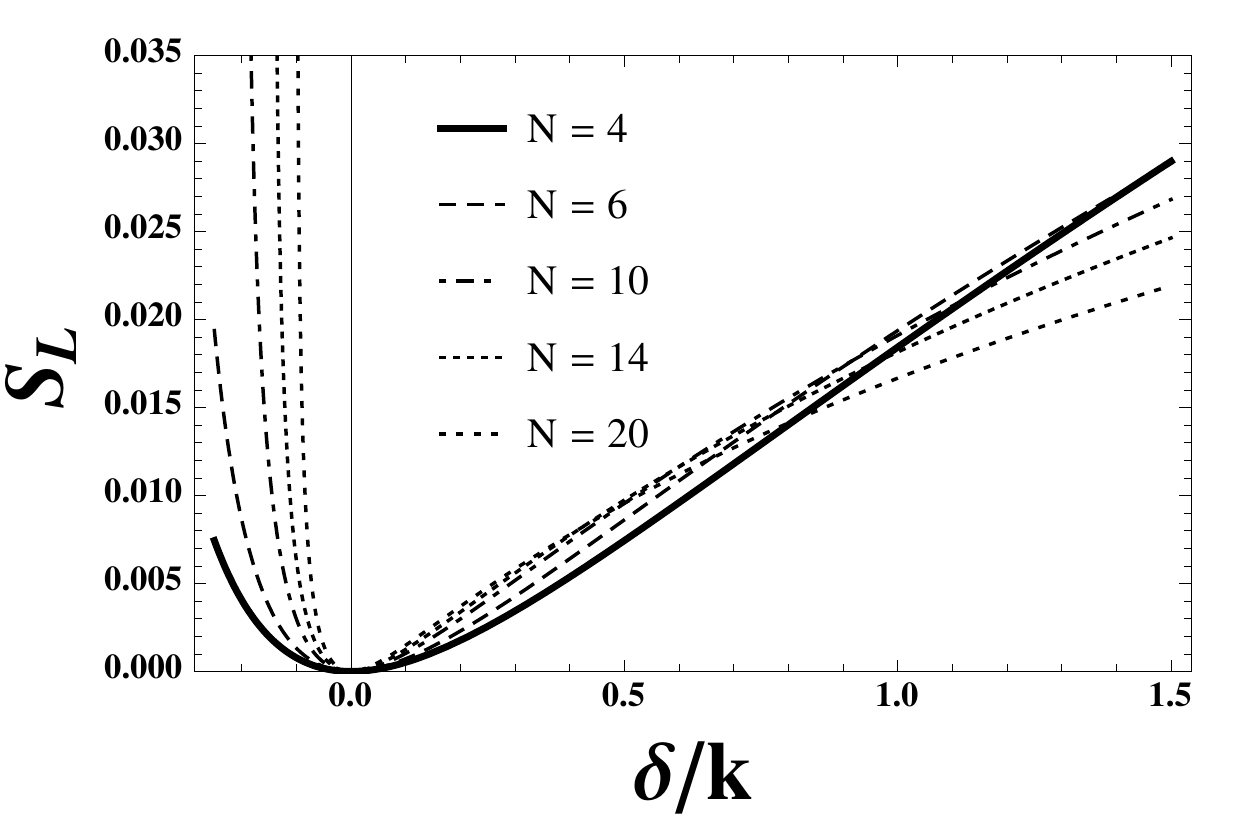} 
 \hfill \caption{Linear entropy of the spinned N-fermion harmonium as a function of the 
 relative coupling constant for various values of $N$. Qualitatively, it is similar to the spinless 
 case in Fig.~\ref{graf:five}.}
 \label{graf:eight}
\end{figure}

The expression \eqref{eq:expre} is similar to \eqref{eq:e7},
except that we have an even number of fermions: $N \to 2N$ and $c_N \to \tilde{c}_{2N}$. 
It is now clear that the conclusions for the spinless fermionic case also hold 
for the full spin case from a qualitative point of view.  Figure  \ref{graf:eight} plots the 
linear entropy for the spinless and the spinned cases as functions of the relative strength. 

\begin{figure}[ht] 
\centering
\begin{subfigure}[b]{.48\textwidth}
 \centering
 \includegraphics[width=7.8cm]{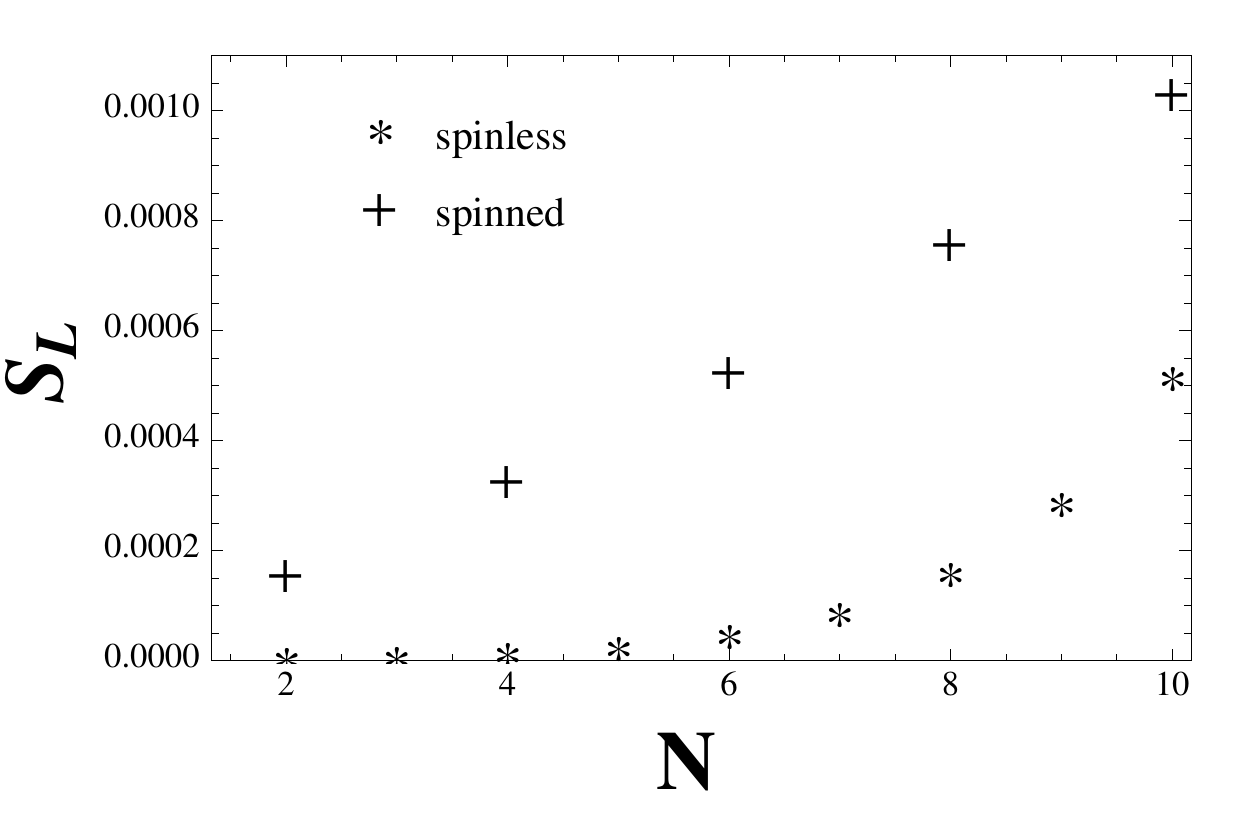} 
\caption{$\dl/k = -1/15$} 
\end{subfigure}
\hfill
\begin{subfigure}[b]{.48\textwidth}
\centering
\includegraphics[width=7.8cm]{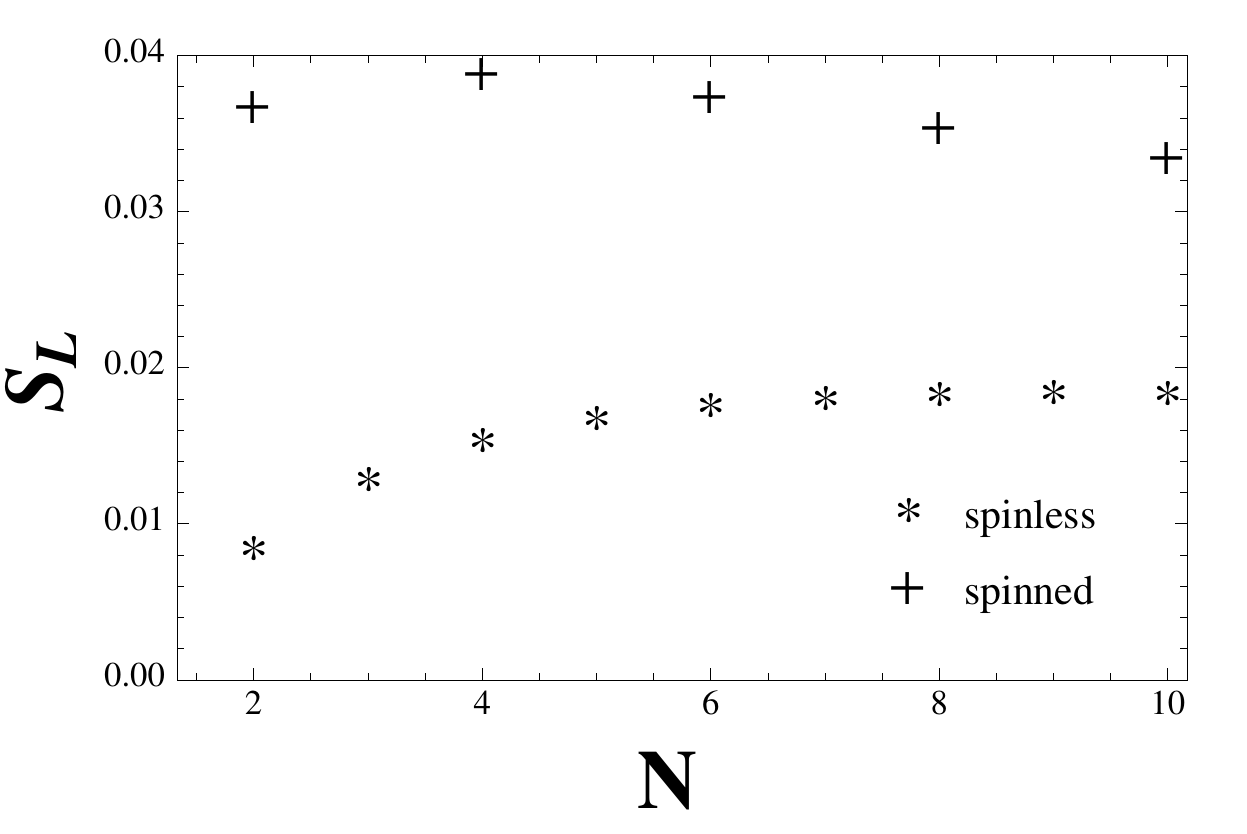}  
\caption{$\dl/k = 2$}
\end{subfigure}
\\
\begin{subfigure}[b]{.48\textwidth}
 \centering
 \includegraphics[width=7.8cm]{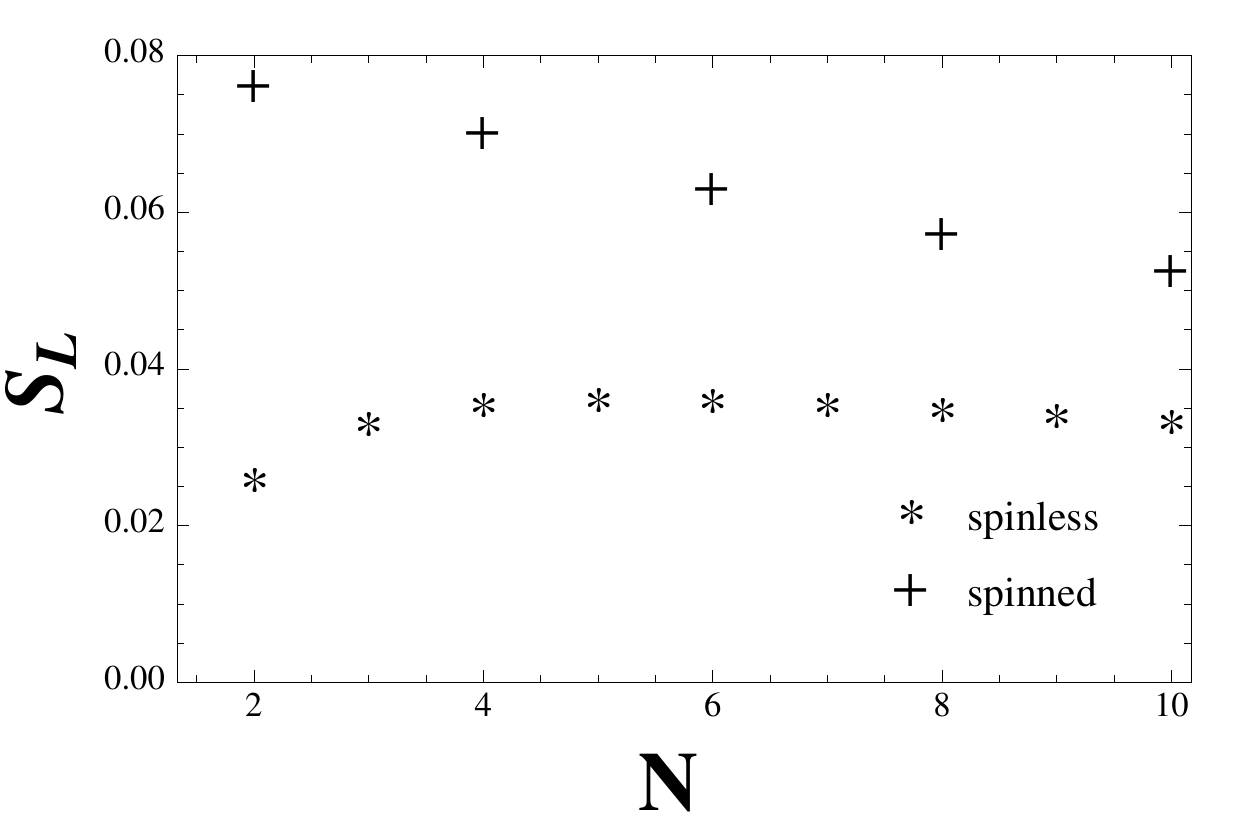} 
\caption{$\dl/k = 4$} 
\end{subfigure}
\hfill
\begin{subfigure}[b]{.48\textwidth}
\centering
\includegraphics[width=7.8cm]{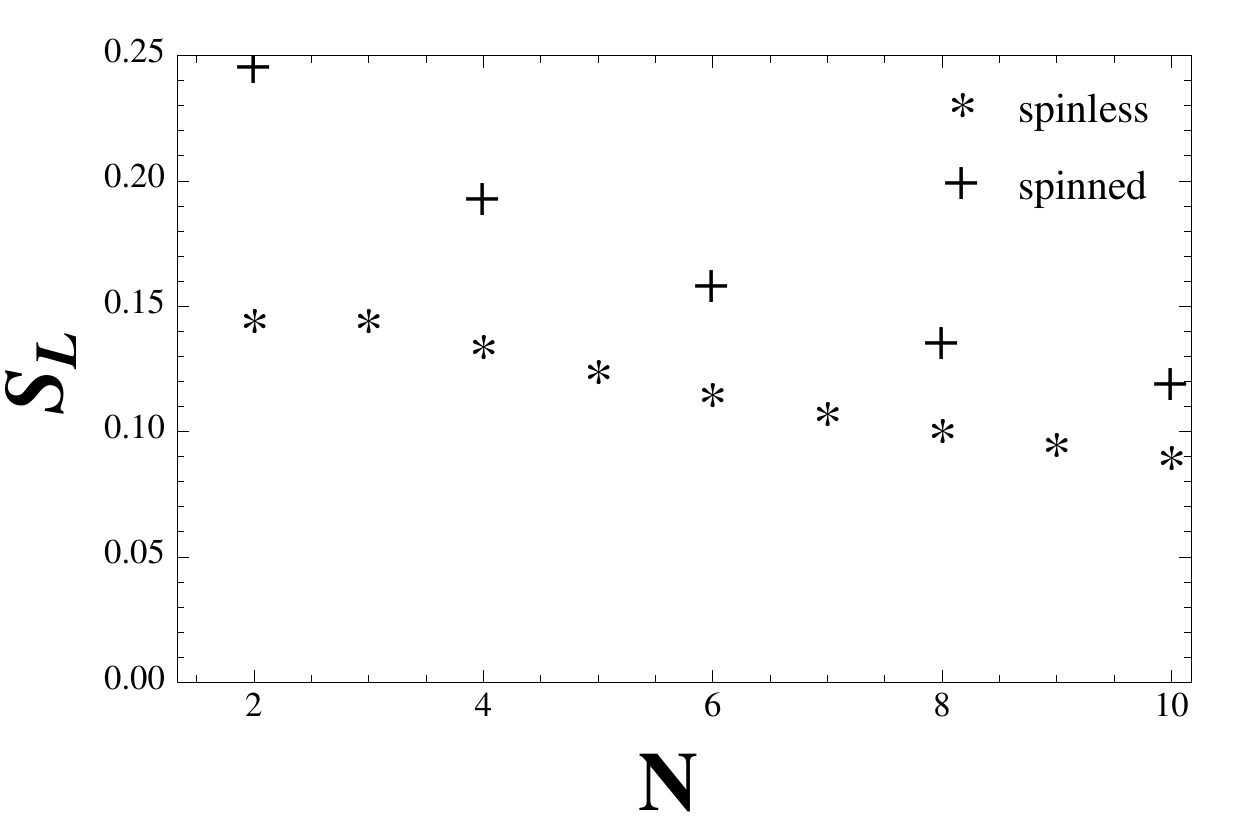}  
\caption{$\dl/k = 22$}
\end{subfigure}
\hfill
\caption{Linear entropy of the one-body reduced density matrix for 
the spinless and spinned N-fermion harmonium as a 
function of the number of particles for different values of $\dl/k$.}
\label{graf:LinearS4}
\end{figure}

For completeness, we have numerically investigated the dependence of 
the linear entropy on the number of particles for some specific values of the 
relative coupling in both spinless and spinned cases (i.e., for spatial and total 
entanglements). Figure \ref{graf:LinearS4} exhibits the results for 
$\dl/k \in \{-1/15,2,4,22\}$. 
In the spinned case the behavior is similar to the spinless system, so that the 
total entanglement behaves similarly as the spatial entanglement. In fact, for 
small values of the coupling the entropy grows when increasing the number of 
particles, while in the strong-coupling regime the situation is the opposite. The 
inclusion of the spin tends to enhance the entropy of the system, being most 
important this phenomenon for small values of the coupling constant. Note 
that in turn the contributions of the spin and the spatial one are of 
comparable size.

\begin{figure}[ht] 
\centering
\includegraphics[width=9cm]{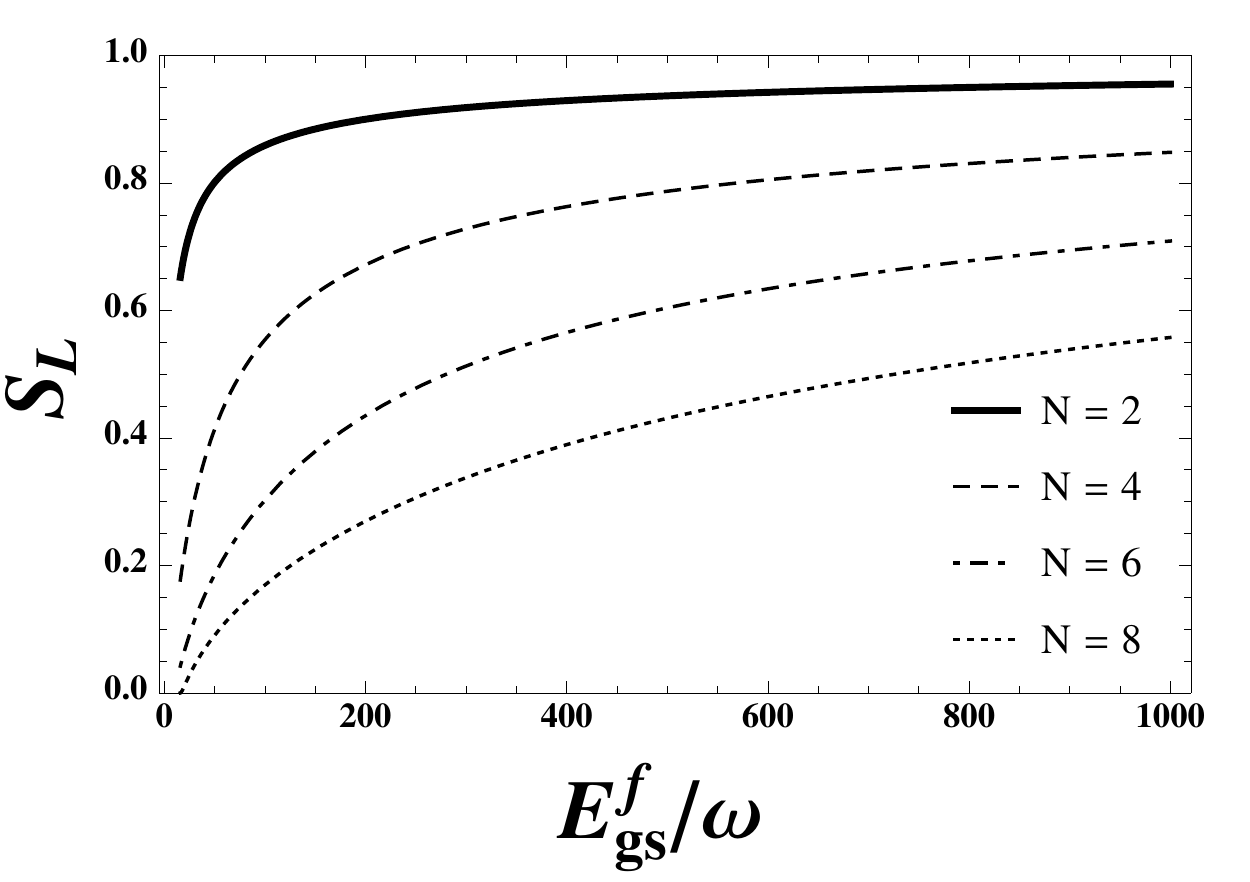}  
\hfill \caption{Linear entropy $S_L$ of the N-fermion spinned harmonium versus the ground-state 
energy $E^f_\gs/\om$ for different values of $N$. The relation between the relative strength
$\dl/k$ and the energy is stated in \eqref{eq:relativestrenghtsp}.}
\label{graf:energyspin}
\end{figure}

Finally, let us investigate the relative behavior of entanglement and energy. 
This is done in Figure \ref{graf:energyspin}, where the linear entropy as a 
function of the ground-state energy is shown for some values of the number 
of particles. Let us recall here that the relation between the relative coupling 
and the energy is given by
\begin{align}
\dl/k = \bigg[\frac{2 E^f_\gs/\om-1}{2 N^2-1} \bigg]^2-1.
\label{eq:relativestrenghtsp}
\end{align}
As in the spinless case, the entropy grows when increasing the ground-state energy. 
For very large values of the dimensionless energy $E_\gs^f/\om$ (i.e., very large 
values of the relative strength $\dl / k$) the gap between the spinless and 
spinned cases decreases.  

% \S 6
\section{Concluding remarks}
\label{sec:conclusion}

In this work we have shown that some entanglement features of 
finite many-particle systems can be understood to a certain extent 
by purely kinematical considerations. This is done by explicitly analyzing 
the entanglement of the N-boson and N-fermion harmonium systems.  
This has been possible because for these harmonic systems we have 
been able to calculate analytically not only the one-body reduced matrix 
for bosons and fermions, but also the von Neumann entropy in the bosonic 
case as well as the linear entropy in the fermionic case. In doing so, we 
complement and extend to harmonic systems with an arbitrary number 
of particles the study of entanglement recently done for various two-electron 
models \cite{HEI, Moshinsky1, NagyP09, YAN, Laetitia, MANZA} as well 
as some helium-like systems \cite{Dehesa, LIN, HUAN, Benenti} and certain 
quantum complex networks \cite{manyos}. 

We have determined the entanglement of these harmonic systems for both 
spatial and spin degrees of freedom analytically in terms of the number of 
particles and the relative interaction strength (or coupling constant). We have 
used the von Neumann entropy and the linear entropy as entanglement quantifiers 
in the bosonic and fermionic systems, respectively. We have found that for positive 
couplings the entanglement of the 3-boson harmonium atom is bigger than the one 
of the 2-boson system, but in general the entanglement of the N-boson harmonium 
decreases when $N$ is increasing. Moreover, the entanglement of a given N-boson 
system grows when the positive coupling constant is increasing; that is, when the 
positive value if the strength of interparticle interaction relative to the confinement well 
is increasing. On the other hand, globally speaking the spatial entanglement of the 
N-fermion harmonium grows when $N$ is increasing for both negative and sufficiently 
small positive values of the coupling constant. Moreover, the entanglement behavior is 
opposite in the positive strong-coupling regime. On the other hand, entanglement of a 
given N-fermion system grows when the coupling constant is increasing. The contribution 
of the spin degree of freedom to the entanglement of the N-fermion system is shown to be 
of positive comparable size to the contribution of the spatial degrees of freedom.

Summarizing, we have shown that in the repulsive case and in the attractive case for 
relatively small values of the coupling constant, the entanglement of both bosonic and 
fermionic N-harmonium atoms grows when the number of particles is increasing, 
basically because purity decreases. However, in the regime of strong coupling the 
situation is exactly the opposite: purity increases (and hence entanglement decreases) 
by adding particles to the system.

% \S Ack
\subsection*{Acknowledgments}

CLBR was supported by a Francisco Jos\' e de Caldas scholarship,
funded by Colciencias. He very much appreciates the warm atmosphere of
the Departamento de F\'{\i}sica At\'omica Molecular y Nuclear,
at the Universidad de Granada. IVT and JSD gratefully acknowledge the MINECO grant FIS2011-24540 and the excellence grant FQM-7276 of the Junta de Andaluc\'ia. In addition, the authors are most grateful to  A.~R. Plastino as well as to J.~L. Alonso, A. Botero, J.~M. 
Gracia-Bond\'ia and J.~C. V\'arilly for helpful and illuminating discussions.

\appendix

\section{Calculation of the integral $\int_{-\infty}^{\infty} e^{-u^{2}} H_{k}(a-cu)H_{k}(b-cu)\, du$}
\label{apendicea}

In this section we calculate the integral
\begin{eqnarray}
\label{eq:e1}
\int_{-\infty}^{\infty} e^{-u^{2}} H_{k}(a-cu)H_{k}(b-cu)\, du ,
\end{eqnarray}
where $a$, $b$ and $c$ are real numbers and $H_k(x)$ is the Hermite polynomial
of degree $k$. We use the following property of the Hermite polynomials
\begin{align*}
\label{eq:e2}
H_{n}(-z) = (-1)^{n}H_{n}(z),
\end{align*}
as well as the power series of these polynomials around $z=z_{0}$
\begin{equation*}
H_{n}(z) = \sum_{k=0}^{n} 2^{k} {n \choose k} H_{n-k}(z_{0}) (z-z_{0})^{k}.
\end{equation*}
This expression allows us to rewrite each Hermite polynomial in the following
form:
\begin{align}
\label{eq:e3}
H_{k}(a-cu) {\underset{z_{0}=-a}{=}}
 (-1)^{k} \sum_{n_{1}=0}^{k} (2c)^{n_{1}} {k \choose n_{1}} H_{k-n_{1}}(-a) u^{n_{1}}.
\end{align}
Inserting \eqref{eq:e3} into \eqref{eq:e1} we obtain
\begin{align}
\label{eq:e8}
&\int_{-\infty}^{\infty} e^{-u^{2}}  H_{k}(a-cu)H_{k}(b-cu)\, du \nonumber \\
& \quad = \int_{-\infty}^{\infty} e^{-u^{2}} (-1)^{k} \sum_{n_{1}=0}^{k}  (2c)^{n_{1}} {k \choose n_{1}} H_{k-n_{1}}(-a) u^{n_{1}} 
\, (-1)^{k}\sum_{n_{2}=0}^{k} (2c)^{n_{2}} {k \choose n_{2}} H_{k-n_{2}}(-b) u^{n_{2}} \nonumber
\\ 
 & \quad = \sum_{\substack{n_{1},n_{2}=0 \\ n_1+n_2 \, {\rm even}}}^{k} 
 (2c)^{n_{1}+n_{2}}  {k \choose n_{1}} {k \choose n_{2}} H_{k-n_{1}}(-a) H_{k-n_{2}}(-b) \, 
  \Gamma \left(\frac{n_{1}+n_{2}+1}{2}\right).
\end{align}

\section{The Gaussian integral $\mathcal{I}_{(n,m)}$}
\label{gaussianas}

In the following, we calculate the integral
\begin{align}
\mathcal{I}_{(n,m)} =  \int^{\infty}_{-\infty} \int^{\infty}_{-\infty} 
x^n y^m e^{-a (x^2+y^2) +2c xy} \, dx \, dy.
\end{align}
Here we perform the change of coordinates
\begin{align}
x = u + v \word{and} y = u - v
\end{align}
and the integral now reads:
\begin{align}
\mathcal{I}_{(n,m)} = 2 \sum^n_{i=0}\sum^m_{j=0}
\binom{n}{i}\binom{m}{j} (-1)^j   \int^{\infty}_{-\infty} \int^{\infty}_{-\infty} u^{n+m-i-j} v^{i+j}
\, e^{-\la_1 u^2 - \la_2 v^2} \, du \, dv,
\end{align}
with $\la_1 = 2(a - c)$ and $\la_2 = 2 (a+c)$. 
From the form of the (double) integral, on using polar 
coordinates in the $(u,v)$-plane, $u = R \cos\phi$ and $v = R \sin\phi$,
since the exponential is an even function of $\phi$, the double integral 
vanishes if $i+j$ is odd. Switching $u$ and $v$ in the polar coordinates
we likewise see that it vanishes if $(m+n)-(i+j)$ is odd.

The integral thus becomes
\begin{align}
& \mathcal{I}_{(n,m)} \nonumber \\
&= 2 \sum^n_{i=0}\sum^m_{j=0}
\binom{n}{i}\binom{m}{j} (-1)^j
\la_1^{-(n+m-i-j+1)/2}\la_2^{-(i+j+1)/2}
 \int^{\infty}_{-\infty} \int^{\infty}_{-\infty} u^{n+m-i-j} v^{i+j}
\, e^{- u^2 - v^2} \, du \, dv \nonumber \\
&=
2 \sum^n_{i=0}\sum^m_{j=0}
\binom{n}{i}\binom{m}{j} (-1)^j  
\la_1^{-(n+m-i-j+1)/2}\la_2^{-(i+j+1)/2} \nonumber \\ & \quad \x
\Ga\Big(\frac{n+m-i-j+1}{2}\Big)\Ga\Big(\frac{n+m+1}{2}\Big)
\frac14 [1+(-1)^{n+m-i-j}] [1+(-1)^{i+j}].
\end{align}

\end{document}